\pdfoutput=1
\documentclass[preprint,12pt,authoryear]{elsarticle}
\makeatletter
\def\ps@pprintTitle{%
 \let\@oddhead\@empty
 \let\@evenhead\@empty
 \def\@oddfoot{\footnotesize\itshape
   Preprint -- Submitted to Spatial Statistics\hfill\today}%
 \let\@evenfoot\@oddfoot}
\makeatother
\usepackage{array}
\usepackage[a4paper, total={7in, 10in}]{geometry}
\usepackage{graphicx}
\usepackage{booktabs}
\usepackage{threeparttable}
\usepackage{natbib}
\usepackage{xurl}
\usepackage{lmodern}
\usepackage[english]{babel}
\usepackage{hyperref}
\usepackage{mathtools}
\usepackage{amsmath}
\usepackage{amssymb}
\usepackage[dvipsnames]{xcolor}
\usepackage[algo2e,ruled,linesnumbered]{algorithm2e}
\usepackage{widetable}
\usepackage{placeins}
\usepackage{tikz}
\usetikzlibrary{arrows.meta, positioning, shapes, calc}

\begin{document}

\begin{frontmatter}

\title{A Top-Down Scale Approach for Multiscale Geographically and Temporally Weighted Regression}

\author[aff1]{Ghislain Geniaux\corref{cor1}}
\ead{ghislain.geniaux@inrae.fr}
\author[aff1]{César Martinez}
\author[aff2]{Samuel Soubeyrand}

\cortext[cor1]{Corresponding author}

\address[aff1]{INRAE, UR 767 Ecodeveloppement, 84914 Avignon, France}
\address[aff2]{INRAE, BioSP, 84914 Avignon, France}

\begin{abstract}
This paper presents \texttt{tds\_mgtwr}, a multiscale geographically and temporally weighted regression (MGTWR) estimator with covariate-specific spatial and temporal bandwidths. Its calibration extends the Top-Down Scale (TDS) scheme of Geniaux (2026) to space--time: a product-separable kernel, restricted searches over ordered geometric grids, and multiscale backfitting with an importance-driven update order and safeguards against degenerate fits. Boundary candidates let each component become spatial-only, temporal-only or global. We also propose out-of-sample prediction for multiscale geographically weighted regression (MGWR) and MGTWR.

Monte Carlo experiments on a deliberately complex design, with spatial and spatio-temporal coefficient patterns under cyclic and non-cyclic temporal settings, show that \texttt{tds\_mgtwr} has the lowest mean coefficient error in every configuration, against single-scale, eigenvector-based and released MGTWR estimators, with an advantage growing with sample size and signal strength. It is also the fastest of the backfitting estimators, which makes large samples with many covariates tractable. In two applications, MGTWR is the most accurate model for house prices, with more than $40{,}000$ sales, although not distinguishable from XGBoost by a paired bootstrap, and, for beet yellows severity, the most accurate model in one of four seasons, indistinguishable from the best in two others and behind a single space--time bandwidth (GTWR) in the remaining one. Its gain is largest when the data are abundant and their space--time structure is complex. The selected scales also tell which effects vary over space, over time or both, and which do not.
\end{abstract}


\end{frontmatter}

\section{Introduction}
Capturing the heterogeneity of relationships between variables across space and time remains a fundamental challenge in quantitative geography, environmental economics, and epidemiology. While global models assume stationarity, recent methodological advances have established robust frameworks for estimating varying effects. We distinguish five broad approaches to space--time varying relationships, which share the goal of modeling non-stationarity but rely on distinct estimation strategies. We briefly review the first four before focusing on kernel-based regression, which motivates the present contribution.

Four families of methods address this heterogeneity through global representations. Eigenvector spatial filtering decomposes map patterns into orthogonal eigenvectors to estimate spatially varying coefficients \citep{griffith2003spatial}, and has been extended to space--time through separable or Kronecker-product eigenvectors \citep{griffith2012space,murakami2019eigenvector}, up to the spatially and temporally varying coefficient (STVC) model of \citet{murakami2025stvc}. Bayesian spatially varying coefficient models treat coefficients as latent processes \citep{gelfand2003spatial}, generalized to dynamic spatio-temporal models with state-space formulations \citep{gamerman2008spatial,lopes2011generalized,banerjee2014hierarchical}. Geo-additive models capture heterogeneity with penalized splines of the coordinates \citep{kammann2003geoadditive,wood2006generalized} and tensor-product smooths in space--time \citep{wood2008fast,augustin2009modeling}, relying on global regularization rather than local weighting. Machine-learning extensions, building on functional gradient boosting \citep{buhlmann2007boosting}, further relax structural assumptions, from geographical random forests \citep{georganos2021geographical} to space--time stacking \citep{hengl2018random}, and can also learn the weighting matrix of a local regression \citep{wu_geographically_2021}.

The fifth approach, kernel-based local regression, is the oldest. It originates in locally weighted regression \citep{mcmillen1996one} and geographically weighted regression (GWR; \citealt{brunsdon1996geographically,fotheringham2002geographically}) for cross-sectional data, which were extended to space--time by integrating time into the kernel metric, giving geographically and temporally weighted regression (GTWR; \citealt{huang2010geographically,wu2014geographically,fotheringham2015geographical}). Several extensions of GTWR followed: a geographically and temporally weighted likelihood for nonlinear specifications such as logistic regression \citep{wrenn2014geographically}, mixed formulations combining constant and varying coefficients \citep{liu2017mixed}, and seasonal kernels for periodic temporal patterns \citep{ge2016construction,du2018extending}. Multiscale GWR (MGWR) then introduced covariate-specific bandwidths \citep{fotheringham2017multiscale}, leading to multiscale GTWR (MGTWR), in which each predictor has its own spatial and temporal scales \citep{wu2019multiscale,zhang2021multiscale,yu2024calibration}. Bandwidths are usually selected by the AICc \citep{hurvich1998smoothing} or by cross-validation \citep{cleveland1979robust}.

Calibrating MGTWR amounts to selecting a spatial and a temporal bandwidth for every covariate, and three strategies have been proposed. The original estimator \citep{wu2019multiscale} extends the backfitting algorithm of MGWR \citep{fotheringham2017multiscale}: starting from a single-scale GTWR fit, each coefficient component is updated in turn by smoothing its partial residual, and its bandwidths are re-selected at each iteration by a local search of the AICc---a golden-section search in the implementation used in recent applications \citep{zhang2024exploring,muniu2026spatial}, the Python package \texttt{mgtwr} \citep{sun2024mgtwr}. This sequential re-selection may require many iterations to converge and does not guarantee convergence to a global optimum \citep{yu2024calibration,geniaux2026top}: evaluated over covariate-specific spatial and temporal bandwidths, the AICc defines a non-convex objective whose geometry may exhibit local irregularities and order-dependent convergence behavior.

\citet{yu2024calibration} therefore replace the local re-selection with a global search: the coefficients are estimated by backfitting at fixed bandwidths, and the combination of spatial and temporal bandwidths is optimized by a genetic algorithm that scores each candidate by the AICc of the complete model. Both backfitting estimators share three limitations. First, they are computationally demanding: every component update, or every candidate combination in the global search, requires many local fits. They have therefore been evaluated only on small balanced panels with few covariates. In the simulations of \citet{yu2024calibration}, these panels have at most $4{,}000$ observations and three coefficients, an intercept and two covariates. Second, the simulated coefficient surfaces on which they have been assessed vary smoothly in space and time---linear or polynomial trends and Gaussian random fields with a squared-exponential covariance \citep{yu2024calibration}---and include no complex spatial or spatio-temporal pattern, such as compact localized features, propagating waves or cyclic temporal variation. Third, neither treats the nested specifications---purely spatial, purely temporal or global components, as in hybrid or mixed models---as explicit candidates: at best, they are reached at a bound of the search interval or approximated by the largest bandwidth, which still down-weights distant observations. The two-step local-linear estimator of \citet{wu2026efficient} removes the computational bottleneck by dropping the iterations. In the first step, all coefficients are estimated by a local-linear GTWR at a common, deliberately undersmoothed spatio-temporal scale. In the second step, each coefficient is re-estimated once, by a local-linear regression of its partial residual---the response minus the first-step contributions of the other covariates---with its own spatial and temporal bandwidths, optimized one after the other. On their simulated surfaces, \citet{wu2026efficient} report an accuracy equal to or slightly better than that of backfitting. On the more complex surfaces of the balanced panel of Section~\ref{sec::comparator_mc}, however, the root mean squared error of its coefficient estimates exceeds that of the local backfitting estimator released by \citet{yu2024calibration} by $8$ to $18\%$, depending on the coefficient.

In this article, we extend the Top-Down Scale (TDS) strategy, originally proposed for MGWR \citep{geniaux2026top}, to the spatio-temporal setting, in order to address these limitations within a single backfitting estimator. TDS replaces an unconstrained search over all bandwidth combinations by a coarse-to-fine descent along ordered grids of candidate scales: at each component update, only a small set of neighboring spatial, temporal and joint moves is evaluated, and the move with the largest reduction of the univariate AICc is retained. Each update thus evaluates a few dozen candidate pairs around the current one, instead of searching the whole bandwidth interval on each axis or the whole grid of combinations. Truncated kernels also restrict each local fit to its neighbors, which keeps backfitting tractable for samples of tens of thousands of observations. The descent and an importance-driven update order are designed to limit unnecessary local recalibrations. Safeguards against degenerate local fits keep the descent away from the over-localized solutions to which sparse sampling and local collinearity can lead: grid floors guarantee a minimal local sample, saturated candidate fits are excluded, and the local systems are solved by a rank-revealing method. Finally, boundary candidates on each axis turn the nested specifications into explicit candidates.

The paper makes four contributions. First, it formulates the TDS calibration for MGTWR, with restricted candidate sets on both axes and boundary candidates that let each component become spatial-only, temporal-only or global, so that these nested specifications are selected rather than imposed (Section~\ref{sec:algo}). Second, it proposes out-of-sample prediction for MGWR and MGTWR by extrapolating the estimated coefficients with sharpened covariate-specific fitted kernels, a feature absent from the released MGTWR implementations (Section~\ref{subsec:prediction}). Third, it evaluates the estimator on a deliberately complex Monte Carlo design, against single-scale, purely spatial and eigenvector-based alternatives and against three released MGTWR estimators, reporting coefficient accuracy, selected bandwidths and single-core computing times (Section~\ref{sec:simuls} and \ref{app:timing}). \texttt{tds\_mgtwr} is the fastest of the backfitting estimators in every timed configuration, and its neighbor-restricted fits keep the memory needed for estimation compatible with samples of tens of thousands of observations. Fourth, two applications, to plant-disease severity and house prices, show when temporal localization pays off and how the selected spatial and temporal scales can be read geographically (Section~\ref{sec:realdata}).

The remainder of the paper is organized as follows. Section~\ref{sec:methods} presents the spatio-temporal kernels, the \texttt{tds\_mgtwr} estimator and the out-of-sample prediction scheme. Section~\ref{sec:simuls} reports the Monte Carlo experiments: the design and the compared estimators, the results, and the comparison with existing MGTWR estimators on a balanced panel. Section~\ref{sec:realdata} presents the two applications, their predictive comparison with local, spectral and machine-learning benchmarks, and the geographical reading of the scales selected for house prices. Section~\ref{sec:conclusion} concludes with the limitations and perspectives of the approach. The appendices detail the prediction kernels (\ref{A_extrapolation}), update-order effects (\ref{C_order}), the numerical controls of the calibration (\ref{B_controls}), the Monte Carlo settings (\ref{B_mc}), the computing times (\ref{app:timing}), the settings of the applications (\ref{B_apps}) and the inference illustration (\ref{B_inf}).

\section{Methods}
\label{sec:methods}

Geographically and Temporally Weighted Regression (GTWR; \citealt{huang2010geographically,fotheringham2015geographical}) and its multiscale extension (MGTWR; \citealt{fotheringham2017multiscale,wu2019multiscale,yu2024calibration}) take the general form:
\begin{equation}
Y_{i} \;=\; \beta_1(u_i,v_i,t_i)\;+\;\sum_{k=2}^{K}\beta_k(u_i,v_i,t_i)\,X_{k i}\;+\;\epsilon_{i},
\qquad i=1,\dots,n\qquad 
\label{eq:gtwr}
\end{equation}
where $Y_{i}$ is the response for observation $i$ at timestamp $t_i$, $(u_i,v_i)$ are spatial coordinates, $X_{k i}$ is the $k$-th predictor ($K$ coefficients in all, the intercept included), $\beta_k$ are local (space--time varying) coefficients, and $\epsilon_i$ is a mean-zero error term. Throughout, the subscript $i$ indexes a \emph{single observation}, i.e., one spatio-temporal sampling point $(u_i,v_i,t_i)$: several observations may share the same timestamp (cross-sectional data collected at common dates) or the same location (repeated measurements), without any consequence for the formulation, since all spatial, temporal, and cyclic distances used below are defined between pairs of observations rather than between distinct sites or dates. Coefficients are estimated via locally weighted least squares using a spatio-temporal kernel centered at $(u_i,v_i,t_i)$.

The fundamental distinction lies in the handling of scale. In GTWR, all coefficients share \emph{two} common bandwidths---one spatial and one temporal---imposing a uniform scale of non-stationarity across all processes. By contrast, MGTWR relaxes this assumption by assigning \emph{covariate-specific} bandwidths (one spatial and one temporal per covariate), calibrated via multiscale backfitting \citep{fotheringham2017multiscale}. This flexibility allows each relationship $\beta_k$ to operate at its own intrinsic spatial and temporal scale.

Our goal is to estimate these varying coefficients while preserving transparent local neighborhoods. Building on the \emph{Top-Down Scale} (TDS) framework for MGWR \citep{geniaux2026top}, we propose a structured optimization strategy that replaces joint high-dimensional bandwidth search with a coordinate-wise descent over ordered grids of candidate scales. Two design choices are pivotal: (i) a separable kernel that calibrates space and time separately and accommodates domain-specific structures (e.g., seasonality), and (ii) a greedy estimation scheme that moves each bandwidth pair along ordered grids under a local AICc criterion.

\subsection{Kernel formulation}
\label{sec:kernel}

The \texttt{tds\_mgtwr} algorithm relies on a spatio-temporal kernel that combines spatial and temporal components. This kernel defines, for each focal location $i$, a set of weights $K_{i,j}^{(st)}$ determining the contribution of observation $j$ to the local estimation at $i$. Formally, let $d_{ij}^{(s)}$ denote the spatial distance and $d_{ij}^{(t)} = |t_i - t_j|$ the temporal distance. The spatio-temporal weight is defined using a multiplicative combination operator\footnote{The \texttt{tds\_mgtwr} implementation also offers an additive and a hybrid additive/multiplicative combination, $\alpha\,K^{(s)}w^{(t)}+(1-\alpha)\,(K^{(s)}+w^{(t)})$, the two kernels being row-normalized before they are combined and the result afterwards.}:
\begin{equation}
K_{i,j}^{(st)}
\;\propto\;
K^{(s)}\!\left(d_{ij}^{(s)}; h^{(s)}\right)
\;\times\;
w_{ij}^{(t)},
\end{equation}
where $K^{(s)}(\cdot; h^{(s)})$ is a standard spatial kernel (e.g., Gaussian or Bi-square) with spatial bandwidth $h^{(s)}$, and $w_{ij}^{(t)}$ is the temporal weighting term. The product downweights separation in either dimension: with Gaussian marginals it imposes no strict support boundary, and with compactly supported marginals its support is the intersection of the spatial and temporal neighborhoods. 
The temporal weight $w_{ij}^{(t)}$ can be specified in two ways:

\paragraph{1. Linear temporal kernel}
Standard distance decay is applied to the time gap: $w_{ij}^{(t)} = K^{(t)}(d_{ij}^{(t)}; h^{(t)})$. This specification supports asymmetric (one-sided) kernels to respect causality (e.g., $w_{ij}^{(t)} = 0$ if $t_j > t_i$).

\paragraph{2. Seasonal (cyclic) temporal kernel}
For periodic phenomena with cycle length \(C\) (e.g., \(C=365\)), we define the cyclic distance $\delta^{(t)}_{ij}= \min\left(|t_i-t_j| \bmod C, C - (|t_i-t_j| \bmod C) \right)$.
The kernel $w^{\mathrm{cyc}}_{ij} = K^{(t)}(\delta_{ij}^{(t)}; h^{(t)})$ then operates on the position within the cycle rather than linear time, enabling the modeling of recurrent seasonal patterns.

\subsection{\texttt{tds\_mgtwr} algorithm}
\label{sec:algo}

The \texttt{tds\_mgtwr} algorithm extends the backfitting procedure of MGTWR by imposing a structured path through the scale space. Each covariate $k$ has a spatial and a temporal bandwidth, $h_k^{(s)}$ and $h_k^{(t)}$, chosen at each update among a few candidates on ordered grids. Algorithm~\ref{alg:tds_mgtwr} gives the structure of the procedure, and Figure~\ref{flow1} (\ref{B_controls}) a schematic overview.

\paragraph{Bandwidth grids}
As in \citet{geniaux2026top}, the candidate bandwidths form strictly decreasing geometric sequences $\mathcal{H}^{(s)}$ and $\mathcal{H}^{(t)}$ of length $M$, so that one step on each axis is a comparable relative change of scale, in line with standard practice in bandwidth selection \citep{loader1999local, lepski1997optimal}. The grids are data-driven. The upper bounds are set by the data extent. In space, the upper bound is the whole sample, or the neighbor cap when one is set. In time, it is the largest temporal gap within the neighborhoods considered, or the half-cycle $C/2$ for cyclic kernels. The maximum value on each axis acts as a sentinel: selecting it removes weighting along that axis, and lifts the neighbor cap on it, rather than evaluating an ordinary finite-bandwidth kernel. Selecting both sentinels yields a global (OLS-like) component, selecting only the temporal one a purely spatial component, and selecting only the spatial one a purely temporal component. The ordered grids thus embed global, spatial-only, temporal-only and fully spatio-temporal specifications in a single selection framework. These boundary choices remain revisable during subsequent sweeps and are model-selection outcomes, not formal tests of stationarity. The lower bounds are practical safeguards against poorly supported local fits: a minimal local sample in space and a low quantile of the temporal gaps in time (Table~\ref{tab:tds_controls}). They do not characterize all identifiable scales, so that selections at a lower bound should be inspected and, where appropriate, checked under an expanded range. \citet{geniaux2026top} recommends $M$ between $20$ and $40$. We use $M=20$ in the simulations and $M=25$ in the applications, and \ref{B_controls} reports a sensitivity analysis of \texttt{tds\_mgtwr} to $M$. An optional pre-selection further restricts the grids covariate by covariate. Before the search, it computes, for each spatial bandwidth, the smallest temporal bandwidth at which the local weighted variance of the covariate exceeds $5\%$ of its global variance at $90\%$ of the locations. Pairs below this floor are discarded, since a covariate that hardly varies within the kernel window cannot identify a local coefficient, and a covariate with no admissible pair stays at the global scale.

\paragraph{Candidate sets and steepest discrete descent}
Initialization starts from the OLS limit, that is, at the top of both grids, where every component is global, and the grids are then explored downwards, from coarse to fine scales. The algorithm proceeds by backfitting: for each covariate $k$, given its current pair $(h_k^{(s)}, h_k^{(t)})$ and its partial residual, it evaluates a restricted set of candidate pairs $\mathcal{C}_k = \mathcal{C}_k^{(s)} \times \mathcal{C}_k^{(t)}$ around the current pair rather than the whole product grid. $\mathcal{C}_k^{(s)}$ contains the current value, its two neighbors in $\mathcal{H}^{(s)}$, the level just below the smallest bandwidth currently active across covariates, and a few coarser upward steps (the quartile levels of $\mathcal{H}^{(s)}$ above the upper neighbor), which let a coefficient return to larger scales in a single move. The highest of these levels is the spatial sentinel, so that a purely temporal fit always remains a candidate. $\mathcal{C}_k^{(t)}$ is built analogously and always contains the temporal sentinel, so that a purely spatial fit, and with both sentinels a global one, also remains a candidate. Since the downward candidates are adjacent grid levels, the finest scale reached by any component descends one level at a time, and a component moves to a finer scale only when this lowers its criterion: broad-scale structure is fitted first, and finer scales are explored only as long as the data support them. Comparing purely spatial, purely temporal and joint moves on an equal footing lets one dimension stabilize at a broad scale once its structural signal is captured, while the other continues towards finer scales. Because partial residuals change during backfitting and non-monotone sweeps are permitted (see below), these updates do not establish convergence to a coordinatewise minimum of a common objective, and no optimality guarantee is claimed. \ref{C_order} discusses the analogy with block coordinate descent and what it does and does not imply for the update order.

\paragraph{Component update}
Each candidate pair is a univariate GTWR fit of the partial residual on $X_k$, scored by its corrected AICc \citep{hurvich1998smoothing}, and the pair with the smallest value is retained:
\begin{equation}
(h_k^{(s)*}, h_k^{(t)*}) = \underset{(h^{(s)}, h^{(t)}) \in \mathcal{C}_k}{\text{argmin}} \;\text{AICc}(k; h^{(s)}, h^{(t)}).
\end{equation}
Candidates are ordered from coarse to fine on both axes, ties go to the coarsest pair, and candidates with an undefined AICc are ignored. A saturation guard excludes near-interpolating local fits. A bandwidth pair so fine that the local fit absorbs more than $86\%$ of the residual variance of the current complete fit (a residual variance below $\exp(-2)$ times that of the complete fit) is treated as quasi-interpolating and discarded. This rule is a ratio of variances, hence free of units and independent of whether inference is requested. If every candidate is excluded, the pair with the smallest AICc is kept and a warning is issued. The guard, whose counts are returned with each fit, never triggered in the simulations or in the applications, where the pre-selection described above already rules out the least supported pairs. Every local system is solved by a column-pivoted QR decomposition, which detects rank deficiency and keeps leverages within $[0,1]$ when the local design is nearly singular. The global AICc, based on the effective number of parameters of \citet{yu2020inference} adapted to the spatio-temporal setting, is optional and serves post-estimation model comparison only.

\paragraph{Stopping rule and returned state}
The sweeps continue until the relative change of the in-sample RMSE between successive sweeps stays below a tolerance $\varepsilon$, or until a patience rule, an oscillation check or an iteration limit stops them (Table~\ref{tab:tds_controls}). Sweeps that raise the RMSE are allowed and the state returned is the terminal one, as in MGWR. Once the search stops, the bandwidths are frozen and a few additional sweeps (at most three, stopped by the same tolerance) refit the components on their partial residuals, so that the returned coefficients are the backfitting fixed point of the selected scales rather than an intermediate iterate.

\begin{algorithm2e}[H]
\caption{\texttt{tds\_mgtwr}: Top-Down Scale calibration for MGTWR}
\label{alg:tds_mgtwr}
\KwIn{$y$; covariates $\{X_k\}_{k=1}^{K}$; coordinates $(u_i,v_i)$ and timestamps $t_i$; strictly decreasing geometric grids $\mathcal H^{(s)}$, $\mathcal H^{(t)}$ of length $M$; kernel specification (spatial kernel; linear or cyclic temporal kernel; multiplicative combination, $\alpha=1$); update-order strategy (cyclic, random, or importance-driven); tolerance $\varepsilon$; number of polishing sweeps $p$; optional candidate pre-selection}
\KwOut{bandwidth pairs $(h_k^{(s)},h_k^{(t)})$ and coefficient surfaces $\hat\beta_k(u,v,t)$, $k=1,\dots,K$}
\textbf{Optional pre-selection}: for each covariate $k$ other than the intercept and each spatial bandwidth, find the smallest temporal bandwidth at which the local weighted variance of $X_k$ exceeds $5\%$ of its global variance at $90\%$ of the locations, and discard the pairs below this floor from the grids of $k$\;
\textbf{Initialization} (OLS limit): $h_k^{(s)}\!\leftarrow\!\max\mathcal H^{(s)}$, $h_k^{(t)}\!\leftarrow\!\max\mathcal H^{(t)}$ and initial component fits $\hat f_k$ for all $k$; $m\leftarrow 0$; compute $\mathrm{RMSE}_0$\;
\Repeat{$\lvert \mathrm{RMSE}_{m}-\mathrm{RMSE}_{m-1}\rvert/\mathrm{RMSE}_{m-1}<\varepsilon$ on three successive sweeps, or a patience, oscillation, bandwidth ping-pong or iteration-limit safeguard is met}{
  $m\leftarrow m+1$; set the update order of covariates (importance-driven: intercept first, then decreasing standard deviation of the fitted contribution of each component at iteration $m-1$)\;
  \For{each covariate $k$ in that order}{
    partial residual: $e_k \leftarrow y-\sum_{l\neq k}\hat f_l$\;
    candidate set: $\mathcal C_k=\mathcal C_k^{(s)}\times\mathcal C_k^{(t)}$ with $\mathcal C_k^{(s)}=\{h_k^{(s)},\,\text{its two neighbors in }\mathcal H^{(s)},\,\text{the level of }\mathcal H^{(s)}\text{ just below }\min_{l}h_l^{(s)}\}$ and $\mathcal C_k^{(t)}$ analogous---plus coarser quartile levels on both axes, up to the sentinel of each axis; the pairs are independent univariate fits\;
    \For{each $(h^{(s)},h^{(t)})\in\mathcal C_k$}{
      univariate GTWR fit of $e_k$ on $X_k$ with kernel $(h^{(s)},h^{(t)})$; local AICc \citep{hurvich1998smoothing}; saturated fits excluded\;
    }
    steepest discrete descent: $(h_k^{(s)},h_k^{(t)})\leftarrow\arg\min_{(h^{(s)},h^{(t)})\in\mathcal C_k}\mathrm{AICc}$; update $\hat f_k$ accordingly\;
    if inference is requested, update the projection matrices of component $k$\;
    \If{a maximal grid value is selected in one dimension}{
      the component collapses to the nested spatial-only, temporal-only, or global (OLS-like) specification---a boundary solution that remains revisable at later sweeps\;
    }
  }
  compute $\mathrm{RMSE}_m$\;
}
\textbf{Polishing}: with all bandwidths fixed, repeat the sweep over covariates (refit of each $\hat f_k$ on its partial residual) at most $p$ times, stopping as soon as the relative change of the RMSE is below $\varepsilon$\;
return the terminal state (bandwidths, coefficients and, with inference, projection matrices)\;
optionally compute the global AICc via the effective number of parameters of \citet{yu2020inference}, for post-estimation model comparison\;
\end{algorithm2e}

\paragraph{Update order}
Because covariates are updated sequentially on a non-convex criterion, the update order may affect the trajectory and the state reached, and thus the allocation of variation across covariates. Three orderings are available. The fixed cyclic order is the one traditionally adopted for MGWR. A random order at each sweep can mitigate systematic ordering effects, although in non-convex block optimization it still converges only to stationary points \citep{razaviyayn2013unified}. The default importance-driven order is a deterministic analogue of Gauss--Southwell rules, which update first the coordinates with the largest marginal impact on the objective \citep{nutini2015coordinate}, although their acceleration results, established for smooth and strongly convex problems, do not carry over to a discrete non-convex AICc. The importance score of component $k$ is the standard deviation over the observations of its fitted contribution, $q_k=\mathrm{sd}_i(\hat\beta_{ik}x_{ik})$, which depends neither on the unit nor on the level of $x_k$ and requires no extra computation. The intercept is always updated first. This score is a heuristic based on the variation of the current fitted component, not a variance decomposition nor an AICc-gradient rule, and it uses only the previous sweep. In our simulations, the order has little importance on average, even with strongly correlated covariates (\ref{C_order}). It may matter much more on real data, where misspecification, omitted variables and measurement errors add sources of bias that different paths can allocate differently across covariates. The importance-driven order, used throughout this article, is kept as the default for its coarse-to-fine logic: fitting broad-scale structure first restricts the competition among correlated predictors at fine scales to residual variation.

\paragraph{Computational cost}
The cost of a calibration can be counted in local fits. One unit is the set of $n$ weighted local regressions run for a given bandwidth pair, each on a single covariate in the backfitting estimators, that is $O(n\,m)$ operations with truncated kernels, $m$ being the number of non-zero weights, and at most $O(n^2)$. Table~\ref{tab:cost_units} gives this count for \texttt{tds\_mgtwr} and for the three released estimators compared in Section~\ref{sec::comparator_mc}. \ref{app:timing} gives the single-core times, which also reflect implementation choices. Two messages stand out. First, \texttt{tds\_mgtwr} is the cheapest of the backfitting estimators. A backfitting estimator of $K$ pairs costs about $S\,K\,c$ units, for $S$ sweeps and $c$ pairs evaluated per update, and in \texttt{tds\_mgtwr} $c=|\mathcal{C}_k|$ is bounded independently of $n$ and of $M$ (about $25$ pairs per update in the central Monte Carlo configuration, $42$ at most), since each update probes only the current pair, its neighbors on each axis and their combinations, and a few coarser steps up to the sentinels. On the balanced panel of Section~\ref{sec::comparator_mc}, a complete calibration costs about $2\times10^{3}$ units, against about $6\times10^{3}$ for the nested golden-section searches of the Python package \texttt{mgtwr} and at least $5\times10^{4}$ ($3\times10^{5}$ with the number of generations actually observed) for the global search of \citet{yu2024calibration}, which scores whole bandwidth combinations by complete backfittings and, its criterion being the AICc of the complete model, updates the $n\times n$ projection matrices at every candidate, whereas \texttt{tds\_mgtwr} updates them only for the selected pair of each update, and only when inference is requested. The $100$ restarts that the authors suggest to assess convergence would multiply the cost of the global search by $100$. Second, MGTWR\_2SCALL \citep{wu2026efficient} does not backfit: it fits all coefficients jointly at a common scale, then refits each once on its partial residual. It needs only $80$ to $370$ fits, with $4K$ and $4$ local-linear terms rather than one, a local fit with $p$ terms costing between one and $p^{2}$ univariate fits (about two in the timings of \ref{app:timing}), and is $1.3$ to $3.5$ times faster than \texttt{tds\_mgtwr}. The price is a much larger coefficient error, because its partial residuals come from the common-scale fit and are never updated. On the panel, the mean coefficient RMSE of MGTWR\_2SCALL with the alternating search, $0.617$, is $21\%$ above that of \texttt{tds\_mgtwr} ($0.508$). It is also above those of every backfitting estimator, including the global genetic search ($0.583$), whose cost is a thousand times larger. With the released driver, it is nearly twice that of \texttt{tds\_mgtwr} ($0.977$; Table~\ref{tab::panel_rmse}). 

\begin{table}[htbp]
\centering
\caption{Cost of one calibration in local fits (one unit: the $n$ weighted local regressions run for one bandwidth pair, each on one covariate---the fit of a partial residual---in the backfitting estimators and in the global search, with $4K$ terms in the first step of MGTWR\_2SCALL and $4$ in its second). $S$: sweeps; $K$: covariates; $c$: pairs evaluated per update; $g_s$, $g_t$: golden-section steps on each axis, with for MGTWR\_2SCALL a second subscript giving the number of terms of the fits they evaluate ($4K$ in the common-scale step, $4$ in the covariate-specific step); $P$: default population of MATLAB's genetic algorithm ($80$ for $K=4$) and $G$ its number of generations ($G\geq51$, the search stopping after $50$ generations without an improvement larger than $10^{-6}$; median $284$ observed); $S'$: sweeps of one backfitting at fixed bandwidths ($3$ to $4$); $r$: passes of the alternating search until the bandwidth pair stabilizes. Panel: $K=4$, Section~\ref{sec::comparator_mc}, counts observed on the balanced panel (three replications for \texttt{mgtwr} and MGTWR\_2SCALL).}
\label{tab:cost_units}
\small
\begin{tabular}{llc}
\toprule
Estimator & Units & Panel \\
\midrule
\texttt{tds\_mgtwr} (this paper) & \begin{tabular}[t]{@{}l@{}}$S\,K\,c$, $S\approx25$,\\ $c$ bounded ($\approx 20$--$25$)\end{tabular} & $\approx 2\times10^{3}$ \\
\addlinespace[4pt]
\texttt{mgtwr} (Python, golden section) & \begin{tabular}[t]{@{}l@{}}$S\,K\,g_s\,g_t$, nested searches,\\ $S=5$, $g_s g_t\approx300$\end{tabular} & $\approx 6\times10^{3}$ \\
\addlinespace[4pt]
\begin{tabular}[t]{@{}l@{}}Global search of\\ \citet{yu2024calibration}\end{tabular} & \begin{tabular}[t]{@{}l@{}}$G\,P\,S'\,K$, with projection\\ matrices at every unit\end{tabular} & \begin{tabular}[t]{@{}c@{}}$\geq 5\times10^{4}$;\\ $\approx 3\times10^{5}$ observed\end{tabular} \\
\addlinespace[4pt]
\begin{tabular}[t]{@{}l@{}}MGTWR\_2SCALL \citep{wu2026efficient},\\ released driver\end{tabular} & \begin{tabular}[t]{@{}l@{}}$(g_{s,4K}+g_{t,4K})+K\,(g_{s,4}+g_{t,4})$,\\ one golden section per axis\end{tabular} & $\approx 80$ \\
\addlinespace[4pt]
MGTWR\_2SCALL, alternating search & \begin{tabular}[t]{@{}l@{}}$r\,[(g_{s,4K}+g_{t,4K})+K\,(g_{s,4}+g_{t,4})]$,\\ $r\approx 3$--$4$ passes\end{tabular} & $\approx 370$ \\
\bottomrule
\end{tabular}
\end{table}

\subsection{Out-of-sample prediction}
\label{subsec:prediction}

In single-scale GWR, prediction usually re-estimates the calibrated model with the prediction locations as focal points. Prediction with MGWR has received little dedicated attention, largely because this target-centered re-estimation does not carry over directly: the backfitting algorithm requires the observed response variable to update partial residuals, which is unknown out-of-sample. Component smoothers built on training partial residuals could still be evaluated at new targets, as \citet{liu2021prediction} do with the hat-matrix decomposition of \citet{yu2020inference}. In that form, however, prediction requires the covariate-specific $n\times n$ projection matrices of the fitted model, that is, full inference, with $O(n^2)$ memory per covariate and up to $O(n^3)$ operations, which becomes prohibitive for the sample sizes targeted here (\ref{app:timing}). We instead extrapolate the learned coefficients, reusing the kernels and bandwidths selected at estimation without any projection matrix: an implementation choice rather than a general impossibility result. As shown in \citet{geniaux2026top}, MGWR coefficients can be extrapolated from very few neighbors: a short-range adaptive (Shepard-type) kernel applied to the nearest neighbors of the prediction location predicts well, without re-optimizing any bandwidth. This can be understood from the scales learned at estimation. The single bandwidth of GWR is a compromise between covariates, so that its coefficient surfaces are too smooth for some and too noisy for others, and any extrapolation inherits these errors. In MGWR, each surface has already been smoothed at the scale of its own covariate, so that the values of a few neighbors are enough to extrapolate it without further smoothing. For MGTWR, prediction is even less discussed. The released MGTWR implementations compared in this paper, namely the estimator of \citet{yu2024calibration}, MGTWR\_2SCALL \citep{wu2026efficient} and the Python package \texttt{mgtwr}, return in-sample coefficients and fitted values only. The scheme proposed below provides the out-of-sample prediction they lack. 

We propose a generic extrapolation scheme for both MGWR and MGTWR that reuses the kernels (and bandwidths) learned during estimation. To concentrate borrowing on the most relevant neighbors while keeping per-covariate scale information, we apply a simple \emph{kernel sharpening} step: see \ref{A_extrapolation}. This focuses mass on the closest neighbors---consistent with the short-range Shepard kernels (e.g., $4$ to $8$ nearest neighbors) found effective for MGWR prediction in \citet{geniaux2026top}---while preserving the relative scaling implied by the learned bandwidths (across variables and, for MGTWR, across space and time). In practice, for MGWR, moving from a minimal Shepard kernel using 4 to 8 neighbors to bandwidth-aware, covariate-specific smoothing with weight sharpening brings only modest gains in predictive accuracy. \ref{B_apps} gives the sharpening exponents used in the applications. For MGTWR, the refinement is motivated by the \emph{relative} spatial-temporal bandwidth ratios, which differ markedly across covariates, and which a single short-range Shepard kernel cannot honor. The per-covariate, bandwidth-aware smoothing---combined with sharpening---preserves those ratios and limits inappropriate temporal or spatial borrowing. Table~\ref{tab::monte_carlo_results} compares estimators under this prediction setting.

\paragraph{Software}
All estimators (\texttt{tds\_mgwr}, \texttt{tds\_mgtwr}) and prediction routines are implemented in the R package \texttt{mgwrsar} (v1.4.4), whose releases are available on CRAN \url{https://cran.r-project.org/web/packages/mgwrsar/index.html}. The calculations presented in this article correspond to version 1.4.4 of the package, except where stated otherwise (\ref{B_mc}, \ref{app:timing}). The script, seeds and output of each result are recorded in the replication repository (Section~\ref{sec:code}). The implementation supports parallel computation.

\section{Monte Carlo simulations}\label{sec:simuls}
Section~\ref{sec:mc_design} describes the simulated point-pattern design and the compared estimators, and Section~\ref{sec::monte_carlo_results} reports their accuracy and the selected bandwidths. Section~\ref{sec::comparator_mc} then compares \texttt{tds\_mgtwr} with the existing MGTWR estimators on a different, balanced-panel design, the only one their released codes accept.
\subsection{Monte Carlo design}\label{sec:mc_design}
To assess the finite-sample performance of \texttt{tds\_mgtwr}, we use a Monte Carlo design inspired by \citet{geniaux2026top}, with an additional temporal dimension. Two of the four coefficients, $\beta_1(u,v,t)$ and $\beta_2(u,v,t)$, vary in both space and time, whereas the other two, $\beta_3(u,v)$ and $\beta_4(u,v)$, vary only in space. This setup tests the ability of the estimators to identify each coefficient's spatial and temporal scales and to recover coefficient surfaces of contrasted geometry.

\paragraph{Data-generating process}
We simulate
\[
Y_{i}=\beta_1(u_i,v_i,t_i)+\beta_2(u_i,v_i,t_i)X_{1i}+\beta_3(u_i,v_i)X_{2i}+\beta_4(u_i,v_i)X_{3i}+\epsilon_i .
\]
The locations $(u_i, v_i)$ are drawn uniformly on the unit square, and each observation receives a calendar day $t_i \in \{1, \dots, 365\}$ and a year index $year_i \in \{1, 2, 3, 4\}$, drawn independently: locations and dates are not crossed, so the design is a spatio-temporal point pattern, not a panel. The covariates are independent, with $X_{1i}\sim\mathcal U[-\sqrt3,\sqrt3]$ and $X_{2i},X_{3i}\sim\mathcal N(0,1)$, all with zero mean and unit variance. The disturbances are independent standard-normal draws $e_i$, rescaled by a common factor defined below. The phase $s_i\in[0,1)$ of observation $i$ drives the temporal patterns. In the cyclic design, $s_i=(t_i-1)/365$ is the position within the annual cycle, so that the patterns repeat identically every year and a periodic temporal kernel can borrow strength across homologous phases. In the non-cyclic design, $s_i=(t^{\mathrm{abs}}_i-1)/1{,}460$, with $t^{\mathrm{abs}}_i=t_i+365\,(year_i-1)$, so that the patterns unfold only once over the four years. Table~\ref{tab::monte_carlo_results} reports both designs side by side, the kernel-based estimators and STVC using periodic kernels or eigenvectors in the first and non-cyclic ones in the second. Since the two DGPs differ, the table assesses performance under each design, not the benefit of a cyclic kernel on a fixed DGP. The simulations cover $n\in\{600, 1{,}000, 5{,}000\}$ and two values of the signal index $\rho\in\{0.7, 0.9\}$, with $1{,}000$ replications per configuration.

\paragraph{Signal index}
The noise scale targets a global, sample-level signal index. For each replication, the systematic component $\eta_i=\beta_1(u_i,v_i,t_i)+\sum_{k=1}^3 X_{ki}\,\beta_{k+1}(u_i,v_i,t_i)$ and the draws $e_i$ are generated, then $Y_i=\eta_i+a\,e_i$, where $a>0$ solves
\[
\rho=\frac{\sum_{i\in\mathcal I}\eta_i^2}{\sum_{i\in\mathcal I}(\eta_i+a\,e_i)^2},
\]
$\mathcal I$ being the $2n$ generated observations (estimation sample and prediction sample, see below). The index $\rho$ is thus an uncentered signal-to-response ratio, in which the spatio-temporal variability of the coefficients is part of the signal, rather than a centered signal-to-noise variance ratio. Because $a$ depends on the complete draw, the disturbances $a\,e_i$ have no variance fixed in advance, and the results are therefore reported by value of $\rho$. Over the replications of Table~\ref{tab::monte_carlo_results}, $a$ averages $3.6$ at $\rho=0.7$ and $1.8$ at $\rho=0.9$ in both designs, for a systematic component of mean $3.0$ and standard deviation $4.7$, so that the centered signal-to-noise variance ratio is about $1.7$ and $6.4$ and the out-of-sample RMSE of the oracle, which predicts with the true coefficients, equals $a$ (\ref{B_mc}): the prediction errors of Table~\ref{tab::monte_carlo_results} are to be read against this floor.

\begin{figure}[htbp]
    \centering
    \begin{minipage}[t]{0.48\linewidth}
        \centering
        \includegraphics[width=\linewidth]{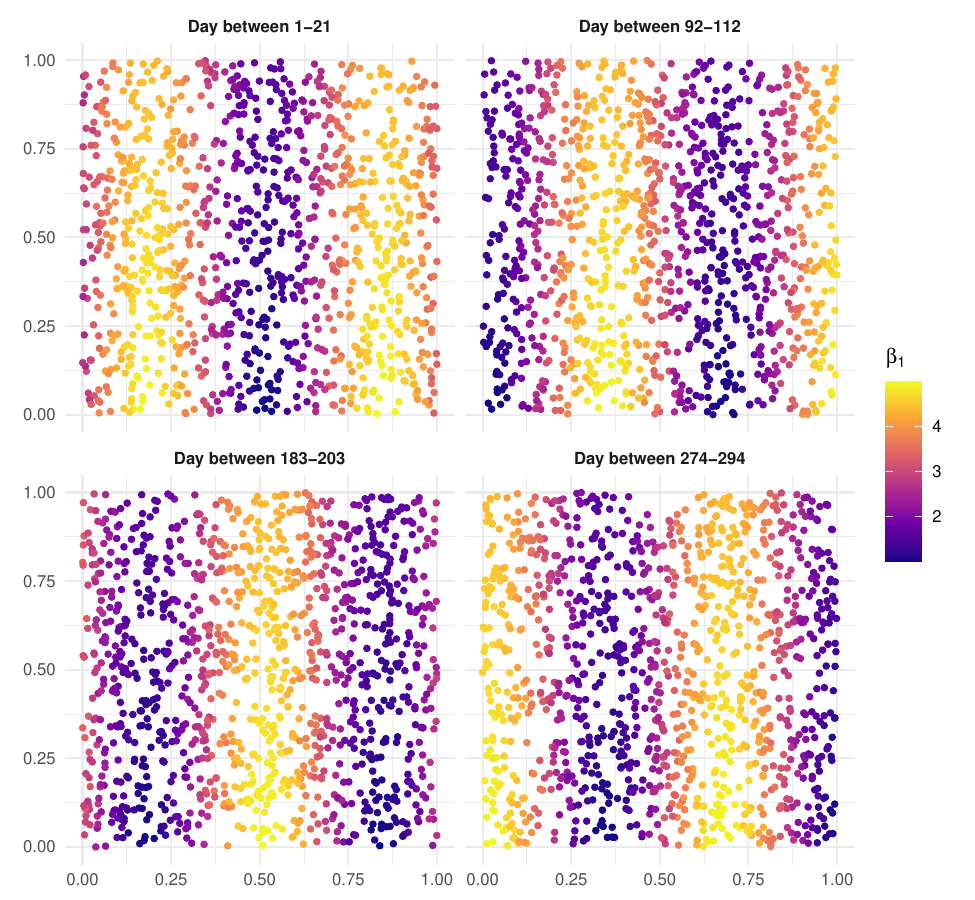}
        \caption{Simulated space-time varying coefficients $\beta_1(u,v,t)$}
        \label{fig::beta1}
    \end{minipage}
    \hfill
    \begin{minipage}[t]{0.48\linewidth}
        \centering
        \includegraphics[width=\linewidth]{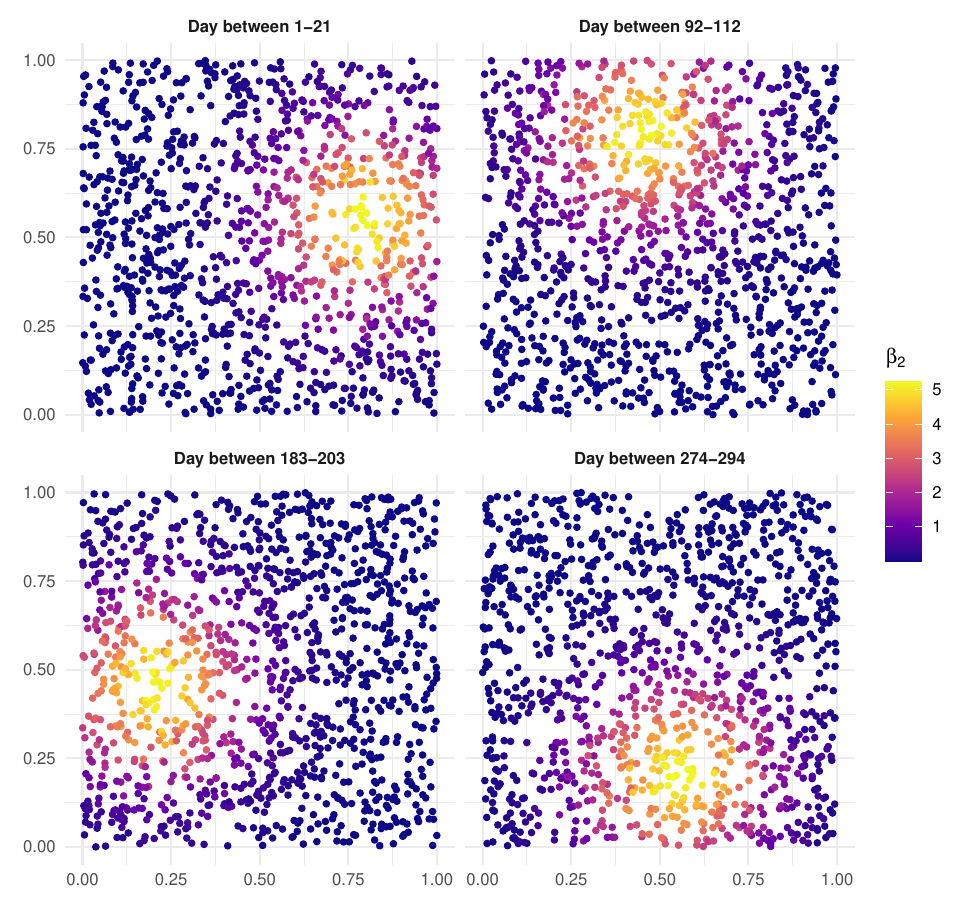}
        \caption{Simulated space-time varying coefficients $\beta_2(u,v,t)$}
        \label{fig::beta2}
    \end{minipage}
\end{figure}

\paragraph{Spatio-temporal coefficients}
Figures~\ref{fig::beta1} and~\ref{fig::beta2} illustrate these two surfaces. The first is a propagating wave whose amplitude decreases with $v$ and whose crests move along $u$ as $s_i$ increases,
\[
\beta_{1}(u_i,v_i,t_i)\;=\;3+2\sin\!\big(\pi(3u_i-2s_i)\big)\,(1-0.3\,v_i);
\]
the second is a Gaussian bump of height $5.25$ whose center completes one rotation around the middle of the study area as $s_i$ runs from $0$ to $1$,
\[
\beta_{2}(u_i,v_i,t_i)\;=\;5.25\,\exp\!\Big(-\frac{\lVert(u_i,v_i)-c(s_i)\rVert^2}{2\cdot0.18^2}\Big),
\qquad
c(s)=\big(0.5+0.3\cos2\pi s,\;0.5+0.3\sin2\pi s\big).
\]
Both surfaces are non-separable in space and time and are not close to a plane at any date. The same two surfaces are used in the balanced-panel comparison of Section~\ref{sec::comparator_mc}.

\begin{figure}[htbp]
    \centering\includegraphics[width=0.9\linewidth]{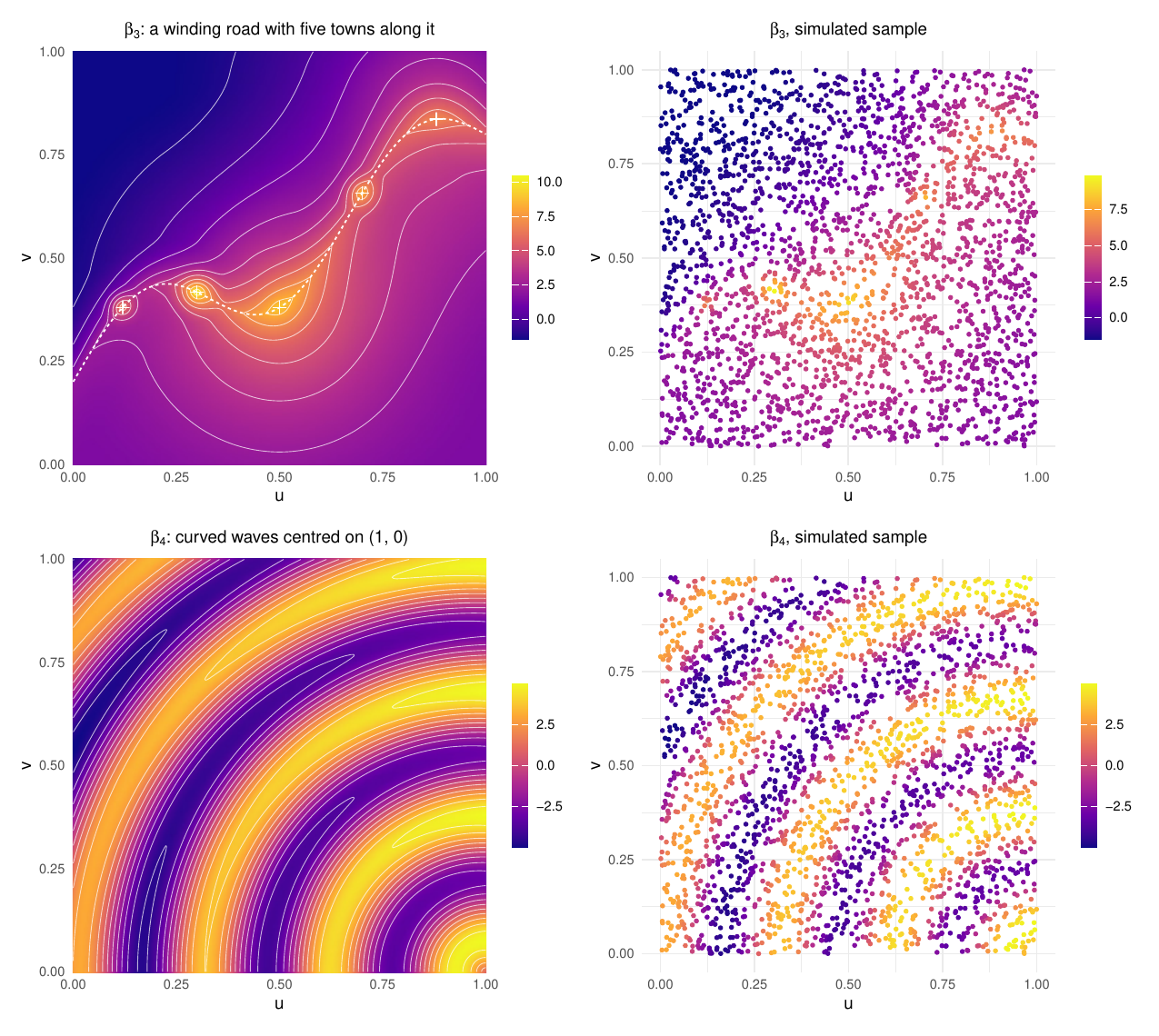}
    \caption{Purely spatial coefficients $\beta_3(u,v)$ (top) and $\beta_4(u,v)$ (bottom). Left: the surface with contour lines; for $\beta_3$, the dashed line is the road and the crosses mark the centers of the five towns. Right: one simulated sample ($n=2{,}500$), in the format of Figures~\ref{fig::beta1} and~\ref{fig::beta2}.}
    \label{fig::bruit}
\end{figure}

\paragraph{Purely spatial coefficients}
The two purely spatial coefficients (Figure~\ref{fig::bruit}) are designed to suit local kernels and global bases to different degrees. $\beta_4$ is a regular wave pattern of constant wavelength, curved around the corner $(1,0)$ and superimposed on a linear trend,
\[
\beta_4(u_i,v_i)\;=\;4\,\sin\!\big(12\sqrt{3}\,\lVert(u_i,v_i)-(1,0)\rVert\big)\;+\;2\,(u_i-0.5),
\]
which a global smooth basis represents efficiently although no single kernel orientation suits the whole surface. $\beta_3$ mimics an accessibility gradient, with features of heterogeneous scale and compact support that local kernels are better placed to follow: a winding road crossing the study area, five towns of different sizes along it, and a background level that differs on either side of the road,
\[
\beta_3(u_i,v_i)\;=\;2.1\,k_{0i}\;+\;\sum_{j=1}^{5}a_j\,k_{ji}\;+\;1.5\,\varsigma_i\prod_{j=0}^{5}\big(1-k_{ji}\big),
\]
where $k_{0i}$ and $k_{ji}$ are compactly supported functions of the distances to the road and to the center of town $j$, $a_j$ the height of town $j$ and $\varsigma_i=\pm1$ the side of the road (road, town positions, radii and heights in \ref{B_mc}). It thus combines a narrow linear feature, three small towns, two large halos and flat areas at two levels. The generating script is included in the replication repository (Section~\ref{sec:code}).

\paragraph{Out-of-sample prediction}
In each replication, $n$ additional observations are drawn from the same DGP and excluded from estimation. Their response is predicted by extrapolating the estimated coefficients. This dimension is restricted to the estimators that come with a prediction method: the released MGTWR implementations of other authors provide none (Section~\ref{subsec:prediction}).

\paragraph{Compared estimators}
The estimators fall into three groups. The first gathers the methods able to recover coefficients that vary jointly in space and time, each at its own scales. Besides \texttt{tds\_mgtwr}, it includes STVC \citep{murakami2025stvc}, which represents each coefficient on spatial, temporal and space--time eigenvectors, and the Python package \texttt{mgtwr} \citep[version 2.0.5;][]{sun2024mgtwr}, which transposes multiscale backfitting to the single space--time distance of \citet{huang2010geographically}. It also includes the MGTWR estimator of \citet{yu2024calibration} and MGTWR\_2SCALL \citep{wu2026efficient}, a two-step local-linear estimator. The second group measures what is lost when a feature of the DGP is ignored: OLS ignores all variation, GWR imposes a single spatial scale and ignores time, MGWR (\texttt{tds\_mgwr}) lets the spatial scale vary across covariates but ignores time, and GTWR accounts for time with a single spatio-temporal scale shared by all coefficients. Each is misspecified by construction, so these estimators are benchmarks for the cost of an ignored feature rather than competitors. The third group consists of the generic learners MARS \citep{friedman1991multivariate} and XGBoost \citep{chen_xgboost_2016}, given the covariates, the coordinates and the date as inputs. They can approximate the DGP for prediction but do not estimate coefficient surfaces, so only their predictive accuracy is reported. Not every estimator enters every experiment: the Python package has no periodic kernel and enters the non-cyclic design only, and the released codes of \citet{yu2024calibration} and \citet{wu2026efficient} accept only a balanced panel and offer neither a periodic kernel nor a prediction method, so that these two estimators are compared with \texttt{tds\_mgtwr} on a balanced-panel adaptation of the design (Section~\ref{sec::comparator_mc}). Kernels, tuning choices and implementation details are given in \ref{B_mc}.

\paragraph{Evaluation criterion}
Since the estimands are coefficient surfaces, the RMSE is defined per coefficient and per replication as an average over the sample points. For replication $r\in\{1,\dots,R\}$ ($R=1{,}000$) and covariate $k$,
\begin{gather*}
\mathrm{RMSE}_k^{(r)}
\;=\;
\sqrt{\frac{1}{n}\sum_{i=1}^{n}
\Big(\hat\beta_k^{(r)}(u_i,v_i,t_i)-\beta_k(u_i,v_i,t_i)\Big)^{2}},
\\
\mathrm{RMSE}_k=\frac{1}{R}\sum_{r=1}^{R}\mathrm{RMSE}_k^{(r)},
\qquad
\overline{\mathrm{RMSE}}=\frac{1}{K}\sum_{k=1}^{K}\mathrm{RMSE}_k ,
\end{gather*}
where $\beta_k(\cdot)$ is the true surface and $\hat\beta_k^{(r)}(\cdot)$ its estimate. Table~\ref{tab::monte_carlo_results} reports $\mathrm{RMSE}_k$ and the mean over the $K$ coefficients. Computing times are not reported in that table, whose estimators span model classes that are not commensurable on that axis. Those of the multiscale spatio-temporal estimators are compared in \ref{app:timing}. Selected bandwidths are reported separately, as distributions, in Figure~\ref{fig:bandwidth_distribution}.
\subsection{Monte Carlo results}
\label{sec::monte_carlo_results}
\begin{table}[!htbp]
\caption{Monte Carlo results with varying signal index $\rho$ and sample size, for the cyclic and the non-cyclic temporal designs ($1{,}000$ replications per configuration). GTWR, \texttt{tds\_mgtwr} and STVC use a periodic temporal kernel or periodic eigenvectors in the cyclic design. The Python package, which has no periodic kernel, enters the non-cyclic design only, and was not run at $n=5{,}000$ because of its computing cost; it offers no prediction method. Best value per column and configuration in bold; bold identifies the smallest reported mean and does not by itself establish a statistically distinguishable difference. $^{\ddagger}$ Mean over the replications with a finite prediction score: a few MGWR predictions at $n=5{,}000$ are non-finite ($992$ and $992$ finite scores out of $1{,}000$ in the cyclic design, $989$ and $995$ in the non-cyclic one, for $\rho=0.7$ and $0.9$); these are failures of the prediction step, not unattempted configurations. \label{tab::monte_carlo_results}}
\centering
\resizebox{\linewidth}{!}{
\begin{tabular}[t]{l cccccc @{\hspace{\dimexpr 2\tabcolsep+0.4cm\relax}} cccccc}
\toprule
 & \multicolumn{6}{c}{Cyclic temporal design} & \multicolumn{6}{c}{Non-cyclic temporal design}\\
\cmidrule(l{3pt}r{3pt}){2-7} \cmidrule(l{3pt}r{3pt}){8-13}
 & \multicolumn{4}{c}{RMSE} & \begin{tabular}[b]{@{}c@{}}mean\\RMSE\end{tabular} & \begin{tabular}[b]{@{}c@{}}Out-of-\\sample\\RMSE\end{tabular} & \multicolumn{4}{c}{RMSE} & \begin{tabular}[b]{@{}c@{}}mean\\RMSE\end{tabular} & \begin{tabular}[b]{@{}c@{}}Out-of-\\sample\\RMSE\end{tabular}\\
\cmidrule(l{3pt}r{3pt}){2-5} \cmidrule(l{3pt}r{3pt}){6-6} \cmidrule(l{3pt}r{3pt}){7-7} \cmidrule(l{3pt}r{3pt}){8-11} \cmidrule(l{3pt}r{3pt}){12-12} \cmidrule(l{3pt}r{3pt}){13-13}
Estimator & $\beta_1(u,v,t)$ & $\beta_2(u,v,t)$ & $\beta_3(u,v)$ & $\beta_4(u,v)$ & $\beta$ & Y & $\beta_1(u,v,t)$ & $\beta_2(u,v,t)$ & $\beta_3(u,v)$ & $\beta_4(u,v)$ & $\beta$ & Y\\
\midrule
\multicolumn{13}{l}{\textbf{Signal index $\rho$ = 0.7, n=600}}\\
\hspace{1em} OLS & 1.226 & 1.364 & 2.187 & 2.895 & 1.918 & 5.453 & 1.226 & 1.362 & 2.187 & 2.896 & 1.918 & 5.449\\
\hspace{1em} GWR (R mgwrsar) & 1.397 & 1.478 & 0.990 & 2.266 & 1.533 & 5.026 & 1.397 & 1.475 & 0.991 & 2.263 & 1.532 & 5.020\\
\hspace{1em} MGWR tds\_mgwr (R mgwrsar) & 1.229 & 1.363 & \textbf{0.932} & 1.738 & 1.315 & 4.622 & 1.229 & 1.360 & \textbf{0.930} & 1.739 & 1.315 & 4.618\\
\hspace{1em} GTWR (R mgwrsar) & 1.267 & 1.345 & 1.005 & 2.369 & 1.496 & 4.941 & 1.351 & 1.418 & 1.012 & 2.318 & 1.525 & 4.945\\
\hspace{1em} MGTWR tds\_mgtwr (R mgwrsar) & \textbf{1.099} & \textbf{1.021} & 0.935 & 1.716 & \textbf{1.193} & \textbf{4.516} & \textbf{1.144} & \textbf{1.059} & 0.936 & 1.723 & \textbf{1.216} & 4.532\\
\hspace{1em} MGTWR (Python mgtwr 2.0.5) &  &  &  &  &  &  & 1.193 & 1.065 & 0.956 & 2.156 & 1.343 & \\
\hspace{1em} STVC (R spmoran) & 1.225 & 1.324 & 0.992 & \textbf{1.574} & 1.279 & 4.517 & 1.224 & 1.222 & 0.989 & \textbf{1.570} & 1.251 & \textbf{4.488}\\
\hspace{1em} MARS u v t (R earth) &  &  &  &  &  & 5.538 &  &  &  &  &  & 5.531\\
\hspace{1em} XGBOOST u v t (R xgboost) &  &  &  &  &  & 5.568 &  &  &  &  &  & 5.565\\
 & & & & & & & & & & & &\\
\multicolumn{13}{l}{\textbf{Signal index $\rho$ = 0.9, n=600}}\\
\hspace{1em} OLS & 1.220 & 1.358 & 2.183 & 2.892 & 1.914 & 4.463 & 1.220 & 1.356 & 2.184 & 2.893 & 1.913 & 4.459\\
\hspace{1em} GWR (R mgwrsar) & 1.319 & 1.406 & 0.856 & 1.895 & 1.369 & 3.713 & 1.318 & 1.403 & 0.857 & 1.894 & 1.368 & 3.707\\
\hspace{1em} MGWR tds\_mgwr (R mgwrsar) & 1.215 & 1.336 & 0.767 & 1.385 & 1.176 & 3.194 & 1.215 & 1.332 & 0.768 & 1.387 & 1.175 & 3.191\\
\hspace{1em} GTWR (R mgwrsar) & 1.173 & 1.278 & 0.857 & 1.988 & 1.324 & 3.557 & 1.286 & 1.370 & 0.859 & 1.916 & 1.358 & 3.558\\
\hspace{1em} MGTWR tds\_mgtwr (R mgwrsar) & \textbf{0.838} & \textbf{0.810} & \textbf{0.738} & 1.392 & \textbf{0.944} & 2.933 & \textbf{0.877} & \textbf{0.832} & \textbf{0.741} & 1.396 & \textbf{0.962} & 2.948\\
\hspace{1em} MGTWR (Python mgtwr 2.0.5) &  &  &  &  &  &  & 1.022 & 0.854 & 0.763 & 1.611 & 1.063 & \\
\hspace{1em} STVC (R spmoran) & 1.145 & 1.080 & 0.759 & \textbf{1.135} & 1.030 & \textbf{2.909} & 1.162 & 0.939 & 0.753 & \textbf{1.121} & 0.994 & \textbf{2.859}\\
\hspace{1em} MARS u v t (R earth) &  &  &  &  &  & 4.446 &  &  &  &  &  & 4.468\\
\hspace{1em} XGBOOST u v t (R xgboost) &  &  &  &  &  & 4.591 &  &  &  &  &  & 4.589\\
 & & & & & & & & & & & &\\
\multicolumn{13}{l}{\textbf{Signal index $\rho$ = 0.7, n=1000}}\\
\hspace{1em} OLS & 1.219 & 1.356 & 2.181 & 2.889 & 1.912 & 5.435 & 1.220 & 1.358 & 2.181 & 2.890 & 1.912 & 5.438\\
\hspace{1em} GWR (R mgwrsar) & 1.366 & 1.444 & 0.880 & 1.814 & 1.376 & 4.810 & 1.367 & 1.444 & 0.879 & 1.816 & 1.376 & 4.814\\
\hspace{1em} MGWR tds\_mgwr (R mgwrsar) & 1.219 & 1.344 & 0.805 & 1.418 & 1.196 & 4.432 & 1.219 & 1.345 & 0.804 & 1.417 & 1.196 & 4.433\\
\hspace{1em} GTWR (R mgwrsar) & 1.209 & 1.309 & 0.882 & 1.914 & 1.329 & 4.672 & 1.329 & 1.407 & 0.882 & 1.845 & 1.366 & 4.690\\
\hspace{1em} MGTWR tds\_mgtwr (R mgwrsar) & \textbf{0.921} & \textbf{0.886} & \textbf{0.795} & 1.400 & \textbf{1.000} & \textbf{4.251} & \textbf{0.962} & \textbf{0.917} & \textbf{0.799} & 1.403 & \textbf{1.020} & \textbf{4.270}\\
\hspace{1em} MGTWR (Python mgtwr 2.0.5) &  &  &  &  &  &  & 1.086 & 0.924 & 0.810 & 1.713 & 1.133 & \\
\hspace{1em} STVC (R spmoran) & 1.209 & 1.294 & 0.902 & \textbf{1.299} & 1.176 & 4.369 & 1.212 & 1.146 & 0.893 & \textbf{1.289} & 1.135 & 4.329\\
\hspace{1em} MARS u v t (R earth) &  &  &  &  &  & 5.405 &  &  &  &  &  & 5.400\\
\hspace{1em} XGBOOST u v t (R xgboost) &  &  &  &  &  & 5.405 &  &  &  &  &  & 5.414\\
 & & & & & & & & & & & &\\
\multicolumn{13}{l}{\textbf{Signal index $\rho$ = 0.9, n=1000}}\\
\hspace{1em} OLS & 1.215 & 1.352 & 2.179 & 2.888 & 1.909 & 4.447 & 1.216 & 1.354 & 2.179 & 2.888 & 1.909 & 4.450\\
\hspace{1em} GWR (R mgwrsar) & 1.285 & 1.369 & 0.741 & 1.490 & 1.221 & 3.414 & 1.286 & 1.369 & 0.740 & 1.492 & 1.222 & 3.419\\
\hspace{1em} MGWR tds\_mgwr (R mgwrsar) & 1.211 & 1.315 & 0.662 & 1.144 & 1.083 & 2.994 & 1.212 & 1.317 & 0.661 & 1.144 & 1.083 & 2.995\\
\hspace{1em} GTWR (R mgwrsar) & 1.040 & 1.168 & 0.752 & 1.625 & 1.146 & 3.218 & 1.218 & 1.306 & 0.746 & 1.550 & 1.205 & 3.263\\
\hspace{1em} MGTWR tds\_mgtwr (R mgwrsar) & \textbf{0.693} & \textbf{0.692} & \textbf{0.617} & 1.077 & \textbf{0.769} & \textbf{2.596} & \textbf{0.716} & \textbf{0.711} & \textbf{0.619} & 1.079 & \textbf{0.781} & \textbf{2.610}\\
\hspace{1em} MGTWR (Python mgtwr 2.0.5) &  &  &  &  &  &  & 0.814 & 0.714 & 0.624 & 1.105 & 0.814 & \\
\hspace{1em} STVC (R spmoran) & 1.064 & 1.052 & 0.673 & \textbf{0.913} & 0.925 & 2.722 & 1.115 & 0.895 & 0.665 & \textbf{0.900} & 0.894 & 2.680\\
\hspace{1em} MARS u v t (R earth) &  &  &  &  &  & 4.307 &  &  &  &  &  & 4.319\\
\hspace{1em} XGBOOST u v t (R xgboost) &  &  &  &  &  & 4.358 &  &  &  &  &  & 4.374\\
 & & & & & & & & & & & &\\
\multicolumn{13}{l}{\textbf{Signal index $\rho$ = 0.7, n=5000}}\\
\hspace{1em} OLS & 1.211 & 1.347 & 2.177 & 2.887 & 1.905 & 5.432 & 1.211 & 1.348 & 2.177 & 2.887 & 1.905 & 5.433\\
\hspace{1em} GWR (R mgwrsar) & 1.274 & 1.347 & 0.564 & 0.953 & 1.035 & 4.329 & 1.275 & 1.347 & 0.565 & 0.952 & 1.035 & 4.329\\
\hspace{1em} MGWR tds\_mgwr (R mgwrsar) & 1.211 & 1.299 & 0.525 & 0.814 & 0.962 & 4.168$^{\ddagger}$ & 1.211 & 1.300 & 0.526 & 0.813 & 0.962 & 4.168$^{\ddagger}$\\
\hspace{1em} GTWR (R mgwrsar) & 0.777 & 0.947 & 0.627 & 1.168 & 0.880 & 4.114 & 0.824 & 0.974 & 0.659 & 1.242 & 0.925 & 4.158\\
\hspace{1em} MGTWR tds\_mgtwr (R mgwrsar) & \textbf{0.581} & \textbf{0.581} & \textbf{0.508} & 0.794 & \textbf{0.616} & \textbf{3.861} & \textbf{0.602} & \textbf{0.599} & \textbf{0.510} & 0.794 & \textbf{0.626} & \textbf{3.865}\\
\hspace{1em} MGTWR (Python mgtwr 2.0.5) &  &  &  &  &  &  &  &  &  &  &  & \\
\hspace{1em} STVC (R spmoran) & 0.974 & 0.998 & 0.529 & \textbf{0.675} & 0.794 & 3.995 & 1.208 & 1.308 & 0.546 & \textbf{0.700} & 0.941 & 4.146\\
\hspace{1em} MARS u v t (R earth) &  &  &  &  &  & 5.182 &  &  &  &  &  & 5.189\\
\hspace{1em} XGBOOST u v t (R xgboost) &  &  &  &  &  & 5.011 &  &  &  &  &  & 5.061\\
 & & & & & & & & & & & &\\
\multicolumn{13}{l}{\textbf{Signal index $\rho$ = 0.9, n=5000}}\\
\hspace{1em} OLS & 1.210 & 1.347 & 2.176 & 2.887 & 1.905 & 4.443 & 1.210 & 1.347 & 2.176 & 2.887 & 1.905 & 4.444\\
\hspace{1em} GWR (R mgwrsar) & 1.229 & 1.306 & 0.452 & 0.744 & 0.933 & 2.843 & 1.229 & 1.306 & 0.453 & 0.743 & 0.933 & 2.843\\
\hspace{1em} MGWR tds\_mgwr (R mgwrsar) & 1.209 & 1.287 & 0.419 & 0.634 & 0.887 & 2.687$^{\ddagger}$ & 1.209 & 1.288 & 0.420 & 0.633 & 0.887 & 2.687$^{\ddagger}$\\
\hspace{1em} GTWR (R mgwrsar) & 0.590 & 0.775 & 0.521 & 0.963 & 0.712 & 2.474 & 0.626 & 0.792 & 0.548 & 1.028 & 0.748 & 2.525\\
\hspace{1em} MGTWR tds\_mgtwr (R mgwrsar) & \textbf{0.424} & \textbf{0.427} & \textbf{0.371} & 0.566 & \textbf{0.447} & \textbf{2.098} & \textbf{0.437} & \textbf{0.441} & \textbf{0.372} & 0.566 & \textbf{0.454} & \textbf{2.103}\\
\hspace{1em} MGTWR (Python mgtwr 2.0.5) &  &  &  &  &  &  &  &  &  &  &  & \\
\hspace{1em} STVC (R spmoran) & 0.783 & 0.789 & 0.382 & \textbf{0.447} & 0.600 & 2.268 & 1.193 & 1.286 & 0.423 & \textbf{0.513} & 0.854 & 2.650\\
\hspace{1em} MARS u v t (R earth) &  &  &  &  &  & 4.115 &  &  &  &  &  & 4.122\\
\hspace{1em} XGBOOST u v t (R xgboost) &  &  &  &  &  & 3.813 &  &  &  &  &  & 3.886\\
\bottomrule
\end{tabular}}
\end{table}

\paragraph{Overall hierarchy}
Table~\ref{tab::monte_carlo_results} shows a stable hierarchy. In the cyclic design, \texttt{tds\_mgtwr} has the lowest mean coefficient RMSE in every configuration, and its margin over the second-best estimator, STVC, grows with the information available, from $7\%$ ($n=600$, $\rho=0.7$) to $26\%$ ($n=5{,}000$, $\rho=0.9$). Among the estimators of coefficient surfaces, OLS has the largest errors, followed by single-scale GWR. MGWR is more accurate than GTWR up to $n=1{,}000$ but not at $n=5{,}000$, where a single temporal scale already captures part of the variation of $\beta_1$ and $\beta_2$. Monte Carlo standard errors of the reported means do not exceed $0.006$ for $\beta_1$--$\beta_3$ and the mean RMSE, $0.019$ for $\beta_4$ and $0.017$ for the out-of-sample RMSE. To compare two estimators, their difference is computed replication by replication, on the same simulated samples. In all $72$ cells of the two designs, the best entry of each column then differs from the second best by more than twice its Monte Carlo standard error, except for two near-ties in the cyclic design at $n=600$, $\rho=0.7$: $\beta_3$ (MGWR and \texttt{tds\_mgtwr}) and the out-of-sample RMSE (\texttt{tds\_mgtwr} and STVC). The paired differences and their standard errors for every cell are included in the replication material.

\paragraph{Spatio-temporal versus purely spatial coefficients}
The advantage of \texttt{tds\_mgtwr} comes first from the two spatio-temporal coefficients, which it recovers best in every configuration: at $n=5{,}000$ and $\rho=0.9$, its error on $\beta_1$ and $\beta_2$ is about one third of that of MGWR and $28$ to $45\%$ below that of cyclic GTWR. A purely spatial smoother recovers little of either surface, since over the year the propagating wave averages out to a constant and the rotating bump to a ring. The two purely spatial coefficients split the kernel and the eigenvector approaches. On the road-and-towns surface $\beta_3$, made of local features of several sizes, \texttt{tds\_mgtwr} is the most accurate estimator, $3$ to $12\%$ below STVC, except in the smallest and noisiest configuration, where it ties with MGWR. On the curved waves $\beta_4$, a regular pattern covering the whole domain, STVC is the most accurate, $7$ to $21\%$ below \texttt{tds\_mgtwr}. An oracle diagnostic (\ref{B_mc}) shows that this gap does not come from the temporal axis or from the bandwidth search: even an isotropic Gaussian kernel fitted at its best bandwidth on the exact partial residual remains above STVC on $\beta_4$, whose eigenvectors represent such a regular global pattern parsimoniously, whereas on $\beta_3$ both the oracle and \texttt{tds\_mgtwr} are below STVC.

\begin{figure}[htbp]
\centering
\includegraphics[width=\linewidth]{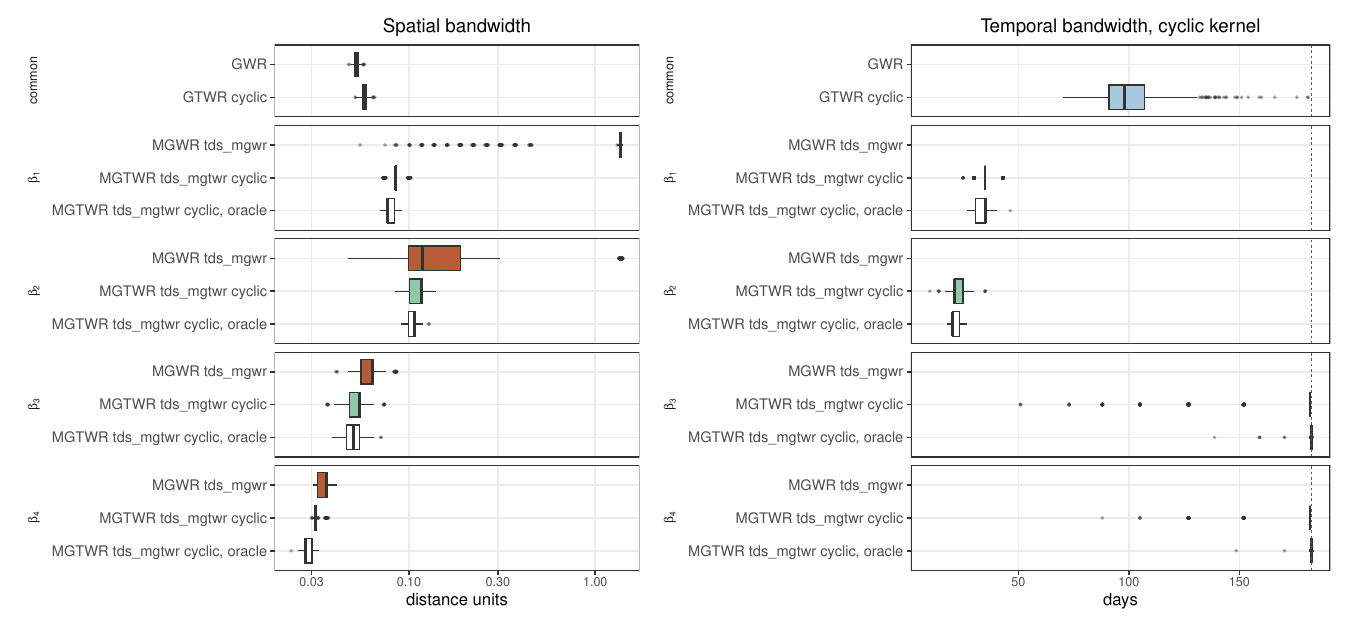}
\caption{Distribution of the selected bandwidths over $1{,}000$ Monte Carlo replications, $n=1{,}000$ and $\rho=0.9$ (cyclic design). Estimators with a single bandwidth (GWR, GTWR) appear once, in the \emph{common} row, their value applying to all covariates; multiscale estimators appear once per covariate. The \emph{oracle} row gives, for \texttt{tds\_mgtwr}, the bandwidth pair minimizing the RMSE of each coefficient fitted on its exact partial residual, computed on the same $1{,}000$ replications. Spatial bandwidths are in distance units (log scale), temporal bandwidths in days; the dashed line marks the admissible maximum of the cyclic kernel (half the cycle, $182.5$ days, rounded to $182$ on the grid, where it is the no-weighting sentinel of Section~\ref{sec:algo}), at which an infinite oracle bandwidth is also plotted. Boxes are quartiles, whiskers $1.5$ IQR.}
\label{fig:bandwidth_distribution}
\end{figure}

\paragraph{Selected bandwidths}
Figure~\ref{fig:bandwidth_distribution} shows the distribution of the selected bandwidths in the central configuration, together with the oracle bandwidths of \texttt{tds\_mgtwr} (the pair minimizing the RMSE of each coefficient fitted on its exact partial residual). Beyond accuracy, the comparison with the oracle row shows how far the selected bandwidths can be read as the scales of the underlying processes. The single-bandwidth estimators apply one spatial scale to all covariates. GWR selects a common spatial bandwidth of $0.052$, close to the oracle value of $\beta_3$ ($0.050$) but about two thirds and one half of those of $\beta_1$ and $\beta_2$ ($0.077$ and $0.107$) and nearly twice that of $\beta_4$ ($0.028$). GTWR adds a common temporal bandwidth of $98$ days, a compromise that corresponds to none of the coefficients: it is about three and five times the oracle values of $\beta_1$ and $\beta_2$ ($35$ and $20$ days), whose seasonal dynamics are therefore oversmoothed, while it imposes a temporal localization on $\beta_3$ and $\beta_4$, which are constant in time. Read as process scales, these selections would suggest that all effects share a common spatial range and a three-month time scale, neither of which exists in the DGP. MGWR, which has no temporal axis, sends $\beta_1$ to a near-global spatial bandwidth in $86\%$ of the replications. Averaged over the year, the propagating wave reduces to its constant mean, so that a coefficient that varies strongly in space and time would be read as stationary. MGWR's bandwidths for $\beta_3$ and $\beta_4$ are also about $30\%$ above the oracle ones. \texttt{tds\_mgtwr} separates the two spatio-temporal surfaces, with temporal bandwidths of about five and three weeks, from the two purely spatial ones, whose temporal bandwidth reaches the admissible maximum in most replications. Its selections lie close to the oracle ones, with little dispersion: within about $10\%$ in space, and $35$ and $21$ days in time against $35$ and $20$ for the oracle. Only the coefficient-specific spatial and temporal bandwidths of MGTWR thus yield smoothing scales commensurate with the oracle scales of each coefficient. Section~\ref{sec:vaucluse_scales} develops this reading on real data.

\paragraph{Non-cyclic design and the Python package}
The hierarchy carries over to the non-cyclic design (right-hand columns of Table~\ref{tab::monte_carlo_results}). With linear time, however, STVC recovers the spatio-temporal coefficients much less well: at $n=5{,}000$ and $\rho=0.7$, its errors on $\beta_1$ and $\beta_2$ remain close to those of OLS, $24$ to $31\%$ higher than with the periodic eigenvectors of the cyclic design. The Python package \texttt{mgtwr}, run up to $n=1{,}000$, is close to \texttt{tds\_mgtwr} on $\beta_2$ and $\beta_3$ but behind on $\beta_1$ and, markedly, on $\beta_4$. Its single space--time distance ties the temporal bandwidth of each coefficient to its spatial one through $\tau$, the ratio that weighs temporal against spatial distance in the space--time distance of \citet{huang2010geographically}, so that a purely spatial coefficient requires $\tau$ at its lower bound, which the search reaches for $\beta_4$ in only $54$ to $89\%$ of the replications. Three of the $4{,}000$ Python runs did not terminate and one STVC fit stopped on an internal error. These runs are left out.

\paragraph{Out-of-sample prediction}
Predictive accuracy follows coefficient accuracy. From $n=1{,}000$ on, \texttt{tds\_mgtwr} is the most accurate predictor in both designs, and its advantage grows with the sample size. At $n=600$, STVC matches or slightly outperforms it (by up to $1\%$ in the cyclic design and $3\%$ in the non-cyclic one). MARS and XGBoost are far behind: up to $n=1{,}000$ neither improves on OLS by more than about $3\%$, and at $n=5{,}000$ the predictive RMSE of \texttt{tds\_mgtwr} is $23$ to $46\%$ below that of XGBoost. Read against the noise floor of Section~\ref{sec:mc_design}, the RMSE of \texttt{tds\_mgtwr} at $n=5{,}000$ exceeds that of the oracle by about $7\%$ at $\rho=0.7$ ($3.86$ against $3.62$) and $14\%$ at $\rho=0.9$ ($2.10$ against $1.84$), whereas that of XGBoost exceeds it by $38$ and $107\%$. The locally linear DGP is one possible explanation of this gap, but the experiment does not separate it from the fixed learner settings (\ref{B_mc}). GWR predictions extrapolate the estimated coefficients with the weights that each target would receive in the local regressions of the training points \citep{geniaux2024speeding}. The other kernel-based estimators use the sharpening parameter $\gamma=8$, the best or near-best value among those recorded on the same test samples ($\gamma\in\{1,2,4,8,12,18\}$). 

\paragraph{Small samples}
In the least informative configuration ($n=600$, $\rho=0.7$), the benefit of modeling spatio-temporal structure is partly offset by estimation variance: MGWR, which selects half as many bandwidths, estimates the purely spatial coefficients about as accurately as \texttt{tds\_mgtwr}. STVC is also attractive in this regime: its eigenvector basis, whose coefficients are estimated as random effects with shrinkage, keeps the effective number of parameters small, so that at $n=600$ it matches or slightly outperforms \texttt{tds\_mgtwr} in prediction, while remaining the most accurate estimator of the regular wave pattern $\beta_4$. When data are sparse or noisy and temporal dynamics are not the focus, simpler or more strongly regularized varying-coefficient specifications may therefore give more stable estimates.

\subsection{Comparison with other existing MGTWR estimators}
\label{sec::comparator_mc}

The results above compare \texttt{tds\_mgtwr} with single-scale and purely spatial multiscale estimators, with the eigenvector-based STVC alternative and, in the non-cyclic design, with the Python package \texttt{mgtwr}. The two other multiscale spatio-temporal estimators with released code, that of \citet{yu2024calibration} in its local (\texttt{mgtwr.m}) and global (\texttt{mgtwrglobal.m}) versions and MGTWR\_2SCALL \citep{wu2026efficient}, accept only a balanced panel and offer no prediction method. They are compared with \texttt{tds\_mgtwr} on a panel adaptation of the design of Section~\ref{sec:simuls}, on coefficient recovery and selected bandwidths.

\paragraph{Panel design and settings} The panel has $N=100$ units on a jittered $10\times10$ lattice observed over $T=20$ periods ($n=2{,}000$), the spatio-temporal surfaces $\beta_1$ and $\beta_2$ of Section~\ref{sec:mc_design} taken in linear time, the purely spatial $\beta_3$ and $\beta_4$ rescaled to the lattice step, $\rho=0.9$ and $1{,}000$ replications. \texttt{tds\_mgtwr} runs in its non-cyclic version with default settings. The local and global versions of the estimator of \citet{yu2024calibration} run through an R port of their released MATLAB files, the global search on the first $100$ replications only because of its cost. MGTWR\_2SCALL runs with its released MATLAB code, once with the released driver and once with the alternating search described by its authors. The design, the kernels and every setting are detailed in \ref{B_mc}.

\begin{table}[!htbp]
\caption{Coefficient RMSE on the balanced panel ($N=100$ lattice units, $T=20$ periods, $1{,}000$ replications); mean over replications, best value per column in bold. $^{a}$ Global genetic search, run on the first $100$ replications only, not compared in bold; on these replications, the mean RMSE is $0.509$ for \texttt{tds\_mgtwr} and $0.550$ for \texttt{mgtwr.m}. \label{tab::panel_rmse}}
\centering
\begin{tabular}[t]{lccccc}
\toprule
Estimator & $\beta_1$ & $\beta_2$ & $\beta_3$ & $\beta_4$ & mean\\
\midrule
\texttt{tds\_mgtwr}, bi-square & \textbf{0.584} & 0.590 & \textbf{0.422} & \textbf{0.435} & \textbf{0.508}\\
Yu--Fotheringham (\texttt{mgtwr.m}) & 0.599 & \textbf{0.584} & 0.473 & 0.542 & 0.550\\
Yu--Fotheringham (\texttt{mgtwrglobal.m})$^{a}$ & 0.638 & 0.621 & 0.509 & 0.565 & 0.583\\
\midrule
MGTWR\_2SCALL, released driver & 0.917 & 0.792 & 0.977 & 1.221 & 0.977\\
MGTWR\_2SCALL, alternating search & 0.646 & 0.640 & 0.559 & 0.621 & 0.617\\
\bottomrule
\end{tabular}
\end{table}

\begin{table}[!htbp]
\caption{Selected bandwidths on the balanced panel: median over replications of the spatial bandwidth, as a site-rank equivalent (the number of units within the bandwidth, the unit itself included), and of the temporal bandwidth in periods. $T$: largest temporal bandwidth, selected in most replications; for \texttt{tds\_mgtwr} (and its oracle) it is the temporal sentinel, i.e., no temporal weighting, whereas for the other estimators it is a finite bi-square window still downweighting distant periods (relative weight $(1-19^2/20^2)^2\approx0.0095$ at a 19-period gap for \citet{yu2024calibration}). These entries therefore do not represent identical temporal weighting. Oracle: for each coefficient, the pair minimizing the RMSE of $\hat\beta_k$ when $\beta_k$ alone is fitted on its exact partial residual, all other coefficients being set to their true values ($20$ replications). $^{a}$ Global genetic search, on the $100$ replications of Table~\ref{tab::panel_rmse}. \label{tab::panel_bw}}
\centering
\begin{tabular}[t]{lcccc}
\toprule
Estimator & $\beta_1$ & $\beta_2$ & $\beta_3$ & $\beta_4$\\
\midrule
\texttt{tds\_mgtwr}, bi-square & 10 / 4 & 17 / 3 & 6 / $T$ & 6 / $T$\\
\texttt{tds\_mgtwr}, bi-square, oracle & 10 / 3.6 & 19 / 2.6 & 6 / $T$ & 6 / $T$\\
Yu--Fotheringham (\texttt{mgtwr.m}) & 10 / 4 & 18.5 / 3 & 6 / $T$ & 6 / $T$\\
Yu--Fotheringham (\texttt{mgtwrglobal.m})$^{a}$ & 13 / 5 & 21 / 3 & 9 / 17 & 6 / 17\\
\midrule
MGTWR\_2SCALL, released driver & 41 / 4.8 & 55 / 3.4 & 36 / $T$ & 24 / $T$\\
MGTWR\_2SCALL, alternating search & 12 / 4.8 & 19 / 3.4 & 7 / $T$ & 6 / $T$\\
\bottomrule
\end{tabular}
\end{table}

\paragraph{Main result} \texttt{tds\_mgtwr} has the lowest mean RMSE (Table~\ref{tab::panel_rmse}). Paired against the local estimator of \citet{yu2024calibration} (\texttt{mgtwr.m}) over the $1{,}000$ replications, it has a lower RMSE on $\beta_1$, $\beta_3$ and $\beta_4$ (by $2.5$, $10.9$ and $19.8\%$, winning $940$, $977$ and $1{,}000$ replications) and a $1.0\%$ higher one on $\beta_2$. The two estimators select nearly the same scales, close to the oracle ones (Table~\ref{tab::panel_bw}). The lower errors on the purely spatial surfaces coincide with a finer resolution of the spatial radius---a distance-valued bandwidth on a geometric grid can place the radius between the orthogonal and diagonal lattice neighbors, where an adaptive integer count cannot---and with a different temporal boundary convention. The comparison evaluates the specified implementations and does not isolate these two contributions.

\paragraph{Global search of Yu and Fotheringham} On the $100$ replications on which it was run, the global genetic search proposed by \citet{yu2024calibration} is, in our implementation-level comparison, both impracticable and less accurate than the local estimator released with it. Its mean RMSE exceeds that of \texttt{mgtwr.m} on the same replications for all four coefficients (by $6.8$, $5.9$, $7.8$ and $4.3\%$, \texttt{mgtwr.m} being more accurate in $94$, $89$, $90$ and $62$ of the $100$ replications), and \texttt{tds\_mgtwr} is more accurate than the global search in $378$ of the $400$ coefficient--replication pairs. It also selects coarser spatial scales than the oracle for $\beta_1$ to $\beta_3$ and, for the purely spatial $\beta_3$ and $\beta_4$, a temporal window of $17$ periods instead of the largest one (Table~\ref{tab::panel_bw}). The genetic algorithm does not reach the optimum of its own criterion: in $75$ of the $100$ replications, the bandwidths selected by the local search give a lower AICc of the complete model (median gap $22.9$). One search took $38$ to $178$ core-hours (median $94$), against about $33$ seconds on one core for \texttt{mgtwr.m}, which rules out the global search for a full Monte Carlo study as well as for routine use. Implementation notes are given in \ref{B_mc}.

\paragraph{Search procedure and accuracy of MGTWR\_2SCALL} The released driver, a single golden-section pass per bandwidth, stops at much larger spatial scales than the alternating search ($24$ to $55$ units against $6$ to $19$) and nearly doubles the RMSE of $\beta_3$ and $\beta_4$. \citet{wu2026efficient} themselves note that the AICc is multimodal. Even with the alternating search, which we implemented as described by \citet{wu2026efficient}, MGTWR\_2SCALL trails the local estimator of \citet{yu2024calibration} on all four coefficients, by $8$ to $18\%$. A plausible explanation is that its step-2 partial residuals come from a first fit at a common scale and are not updated by backfitting. The local-linear boundary behavior and the first-stage scaling of \citet{wu2026efficient} may also contribute, and we did not isolate these factors. The \texttt{tds\_mgtwr} row uses the package defaults, whereas the better MGTWR\_2SCALL row relies on a search that is described by \citet{wu2026efficient} but absent from their released code, and that we had to implement: a user of the released code obtains the first MGTWR\_2SCALL row.

\paragraph{Computing times} \ref{app:timing} compares the single-core computing times of these estimators, of the Python package and of STVC on the same balanced panel, for $n$ from $600$ to $5{,}000$ and $K$ from $2$ to $16$ (Table~\ref{tab:timing_grid}). \texttt{tds\_mgtwr} is the fastest of the backfitting estimators in every configuration---the local estimator of \citet{yu2024calibration} is $4$ to $32$ times slower and their global search $190$ to $780$ times slower where it stays below an eight-hour cap---while MGTWR\_2SCALL, which does not backfit, is $1.3$ to $3.5$ times faster, at the cost of the accuracy losses reported above.

\section{Real-data performance}\label{sec:realdata}

We present two real-data case studies that evaluate \texttt{tds\_mgtwr} in terms of out-of-sample accuracy and interpretability. Both involve multi-year, geocoded observations (Table~\ref{tabdata}) but span contrasting regimes: within-season prediction of plant-disease severity under uneven temporal coverage, and housing-market dynamics with gradual temporal drift over fifteen years.

We compare \texttt{tds\_mgtwr} with OLS, MARS, XGBoost and the two other spatio-temporal estimators, STVC and GTWR, under a common holdout design. For house prices, models are trained on transactions up to 2021 and evaluated on 2022 sales. For beet yellows severity, the late-season observations of each focal year are predicted from its early-season observations and from those of all other years. Accuracy is measured by the out-of-sample RMSE and MAE. In both applications, GTWR and MGTWR combine an adaptive Gaussian spatial kernel with a non-cyclic Gaussian temporal kernel with a fixed bandwidth, the observations being irregularly spaced in time. The seasonality of beet yellows enters through the covariates (day of year) and the training windows, not through a periodic kernel. Specifications were harmonized as far as each model class permits, the locally weighted models combining AICc-driven selection with explicit safeguards against degenerate local fits and severe local multicollinearity. All estimator-specific choices are reported in \ref{B_apps}. The released MGTWR estimators compared in Section~\ref{sec:simuls} do not reappear here: none of them provides an out-of-sample prediction method (Section~\ref{subsec:prediction}), and two of them require a balanced panel, which neither dataset is.

\subsection{Datasets}\label{subsec:datasets}

\paragraph{Beet yellows severity dataset (France)}
This dataset comprises field observations collected across the French sugar-beet production region by technicians of the sugar beet sector and curated by the French Beet Technical Institute (ITB), spanning 2019--2023, and is described in detail by \citet{chauvin2025factor} and \citet{martinez2026}. The observation unit is a field visit: the observed severity $Y$ is a visual assessment, at the field scale, of the proportion of leaf cover affected by virus yellows, and a field may be visited several times within a season. The exercise evaluates within-season prediction of late severity over the summer--autumn window (day of year 201--230) of each season 2020--2023: the training sample contains the observations of the focal season dated before the window together with all observations of the other seasons, earlier and later, and only the late-season observations of the focal season are withheld (Table~\ref{tabdata}). All models are fitted to the logit of the clipped proportion, $\log\{p/(1-p)\}$ with $p=\min\{0.975,\max(0.025,Y)\}$, so that the response is unbounded and Gaussian local regression applies. Predictions are mapped back by the inverse logit, which keeps them within $(0,1)$, and the errors of Table~\ref{tab:beet_all} are computed on this proportion scale, against the observed severities. Guided by the two studies above, $15$ candidate covariates are retained, in seven families: meteorology (degree days, mean and maximum temperatures, precipitation, specific humidity, potential evapotranspiration), land cover (share of grassland), management (seed treatment), landscape context (distance to sugar-beet seed-production fields), biotic pressure (aphid incidence at the visit), two temporal indices (day of year and continuous time) and two regional indicators of past severity in the same season. Every covariate of a withheld observation is computed from information dated before it or recorded at its visit, so that the aphid percentage is a contemporaneous covariate rather than a forecast input. Their names, definitions and time windows are given in Table~S1 of the Supplementary Material.

\paragraph{Vaucluse House Prices (France)}
The \texttt{Vaucluse House Prices} dataset is derived from France's open real-estate transactions database, DVF (\url{https://www.data.gouv.fr/fr/datasets/demandes-de-valeurs-foncieres}). It comprises 44{,}236 sales of detached houses (living area $<300$\,m$^2$ and parcel area between $100$ and $5{,}000$\,m$^2$) in the Vaucluse department between January 2007 and December 2022, geocoded at the parcel level and dated by month. The response is the logarithm of the sale price. Eighteen candidate covariates are used. Four describe time: the year, month and quarter of the sale and a continuous time index. Six describe size: living area and parcel area, each with its squared and cubic terms. The others are the number of swimming pools, the area of outbuildings, the neighborhood-level share of dwellings built before 1945, and the distances to the nearest post office, primary school, hospital, maternity unit and public transport stop.

\begin{table}[htbp]
\caption{Real datasets. Early-season severity $>5\%$: share of the in-sample observations of the target season (up to day $200$) with a beet yellows severity above $5\%$.}
\centering
\label{tabdata}
\resizebox{\textwidth}{!}{%
\begin{tabular}[t]{lllcc}
Dataset Name  & in-sample size & out-of-sample size  & Early-season severity $>5\%$ & Number of covariates \\
\hline
 Vaucluse House Prices  &  40598 (before 2022)  & 3638 (2022) & -- & 18 \\
 Beet Yellows Severity   & 2667 (including 558 in early  2020) & 282 (late 2020) & 50\% & 15\\
 Beet Yellows Severity  & 3030 (including 129 in early  2021)& 179 (late 2021) & 7\% & 15\\
 Beet Yellows Severity  & 2914 (including 236 in early  2022) & 192 (late 2022) & 6\% & 15\\
  Beet Yellows Severity  & 3001 (including 417 in early  2023) &  266 (late 2023) & 5\% & 15\\
\end{tabular}}
\end{table}

\paragraph{Data availability}
The Vaucluse House Prices (France) dataset is provided to the reviewers with the replication material and will be publicly accessible upon publication through the replication repository of this article (Section~\ref{sec:code}):
\url{https://github.com/ggeniaux/tds_mgtwr_replication/}.

In contrast, the beet yellows severity dataset cannot be made publicly available due to confidentiality and anonymity constraints. Its variable dictionary is given in Table~S1 of the Supplementary Material and the analysis specification in \ref{B_apps}, and the companion paper \citep{martinez2026} documents the data in detail.

\subsection{Results}\label{sec:pred_ac}

\subsubsection{Results for beet yellows severity}

\paragraph{Information content of the seasons}
The comparison bears on complete pipelines: covariates are selected separately for each estimator and each season, with the same stepwise rules and from the same candidate set, by the information criterion of the estimator itself (the AICc of GTWR, that of \texttt{tds\_mgtwr}, the AIC of \texttt{spmoran} for STVC, \ref{B_apps}), so that the specifications differ across estimators and seasons (Table~\ref{tab:covars_2020}). Table~\ref{tab:beet_all} shows that no model is the most accurate in every season, and that the ranking of the models depends on how much of the epidemic the focal season documents before the prediction window. The hierarchy is clearest in 2020, the only season whose early-season sample already documents the epidemic: half of its $558$ early-season observations have a severity above $5\%$, against $5$ to $7\%$ of the $129$, $236$ and $417$ observations of 2021--2023 (Table~\ref{tabdata}). The size of the early-season sample is therefore not sufficient: 2023 is nearly as well sampled as 2020, but its early-season plots are almost all healthy and say little about the course of the disease in the target season. In 2020, MGTWR is the most accurate model on both metrics, with an RMSE $12\%$ below that of STVC and $17\%$ below that of GTWR, and the paired bootstrap separates it from every other model, except STVC in RMSE. In 2021, STVC has the lowest RMSE, with OLS, MARS, XGBoost and MGTWR within $2\%$ and GTWR $30\%$ above, yet none of them is distinguishable from the best on the smallest test sample ($179$ observations in $116$ fields). XGBoost has the lowest MAE, with MGTWR within $5\%$. In 2022, XGBoost and GTWR share the lowest RMSE and GTWR has the lowest MAE, STVC and MGTWR lying within $5$ to $7\%$ and indistinguishable from the best. In 2023, GTWR is alone in front: its RMSE is $15\%$ below that of MGTWR, the second best, and its MAE $20\%$ below that of MARS, the second best on this metric. The machine-learning models, which receive the coordinates and time as inputs, are thus competitive in the two seasons whose early sample carries little information on the epidemic (2021 and 2022) and behind the best kernel-based model in 2020 and 2023.

\paragraph{Single and covariate-specific space--time scales}
The three spatio-temporal estimators differ in how they localize in time: GTWR applies one space--time kernel to every covariate, MGTWR gives each covariate its own spatial and temporal scales, and STVC represents the variation of each coefficient on a basis of spatio-temporal eigenvectors. Covariate-specific scales pay off in the two seasons whose training information on the focal season is either the richest or the scarcest: MGTWR improves on GTWR by $17\%$ in RMSE in 2020 and by $22\%$ in 2021, and on STVC by $12\%$ in 2020. A single space--time scale seems to win when the late-season signal is best predicted from the most recent observations: GTWR improves on MGTWR by $5\%$ in 2022 and by $15\%$ in 2023, with a common temporal bandwidth of $196$ and $193$ days, whereas the covariate-specific bandwidths of MGTWR range from about three weeks to the whole series (Table~\ref{tab:covars_2020}). STVC lies between the two in 2020 and stays within $4\%$ of MGTWR from 2021 to 2023. The paired bootstrap tempers these contrasts. In RMSE, the difference between MGTWR and GTWR is distinguishable from zero in 2020 and 2023 only, and none of the three estimators is distinguishable from the best model in 2021 and 2022. In MAE, the difference is also distinguishable in 2021. The drivers of disease severity---weather, landscape, vector abundance---thus vary over the season as well as in space, but the benefit of letting their scales differ from one covariate to another depends on the season, and the sampling of the focal season does not explain the pattern on its own: 2021, the least sampled season, is where MGTWR gains most over GTWR. A plausible reason is that, when the early-season sample of the focal year is small and almost disease-free, covariate-specific temporal scales let the stable effects borrow strength from the other seasons through wide temporal windows---several components receive no temporal weighting in 2021 (Table~\ref{tab:covars_2020})---while a few components remain local in time, whereas the single $178$-day window of GTWR confines every effect to the sparse recent observations.

\paragraph{Specification and inference}
Two methodological points, documented in \ref{B_apps}, qualify these results. First, relying solely on AICc can lead to severe local multicollinearity, a known issue in GWR-type models \citep{wheeler2005multicollinearity} that multiscale formulations reduce without removing \citep{kang2025scale}. The specifications therefore combine AICc selection with explicit local collinearity controls. Second, on the 2020 training sample ($n = 2{,}667$), we computed exact standard errors following \citet{yu2020inference}, which are derived from the covariate-specific projection matrices of the backfitting, and compared them with a local approximation that takes, for each component, the variance of its final univariate GWR fit, ignoring the dependence that backfitting creates between components. On this example, the univariate approximations yield standard errors up to five times smaller than the exact ones and flag a larger share of locations as significant. The comparison and its limits are detailed in \ref{B_inf}.
Exact inference is computationally costly at the size of the Vaucluse House Prices dataset ($n>40{,}000$). We therefore restrict that analysis to predictive performance and coefficient magnitude, and present no significance map, which would rest on the local approximation alone.

\begin{table}[htbp]
\caption{\label{tab:beet_all}Out-of-sample performance on Beet Yellows Severity (2020--2023). The models are fitted on the logit of the severity clipped to $[0.025, 0.975]$; RMSE and MAE are computed on the proportion scale, between the observed severities and the back-transformed predictions. XGBoost: mean over $20$ seeds, tree depth chosen within the training sample (\ref{B_apps}).}
\resizebox{1\linewidth}{!}{
\centering
\begin{tabular}[t]{lcccccccc}
\toprule
\multicolumn{1}{c}{} & \multicolumn{2}{c}{2020} & \multicolumn{2}{c}{2021} & \multicolumn{2}{c}{2022} & \multicolumn{2}{c}{2023} \\
\cmidrule(l{3pt}r{3pt}){2-3} 
\cmidrule(l{3pt}r{3pt}){4-5} 
\cmidrule(l{3pt}r{3pt}){6-7} 
\cmidrule(l{3pt}r{3pt}){8-9}
Model & RMSE & MAE & RMSE & MAE & RMSE & MAE & RMSE & MAE\\
\midrule
OLS & 0.3406 & 0.2774 & 0.1778$^{\dagger}$ & 0.0671 & 0.0826 & 0.0505 & 0.1514 & 0.0776\\
MARS  u v t (R \texttt{earth}) & 0.2563 & 0.2066 & 0.1803$^{\dagger}$ & 0.0704 & 0.0751 & 0.0330 & 0.1756 & 0.0644\\
XGBOOST  u v t (R \texttt{xgboost}) & 0.2213 & 0.1585 & 0.1801$^{\dagger}$ & \textbf{0.0637} & \textbf{0.0582} & 0.0253$^{\dagger}$ & 0.1445 & 0.0732\\
\midrule
STVC (R \texttt{spmoran}) & 0.1985$^{\dagger}$ & 0.1364 & \textbf{0.1773} & 0.0676 & 0.0612$^{\dagger}$ & 0.0253$^{\dagger}$ & 0.1485 & 0.0664\\
GTWR (R \texttt{mgwrsar}) & 0.2104 & 0.1273 & 0.2310$^{\dagger}$ & 0.1095 & 0.0583$^{\dagger}$ & \textbf{0.0243} & \textbf{0.1215} & \textbf{0.0516}\\
MGTWR (tds\_mgtwr, R \texttt{mgwrsar}) & \textbf{0.1746} & \textbf{0.1066} & 0.1807$^{\dagger}$ & 0.0668$^{\dagger}$ & 0.0611$^{\dagger}$ & 0.0260$^{\dagger}$ & 0.1432 & 0.0661\\
\bottomrule
\end{tabular}
}

\vspace{0.4em}
\footnotesize\emph{Notes}: STVC covariates are selected by a forward--backward stepwise procedure on the AIC returned by \texttt{spmoran}, starting from all covariates. The kernel-based models extrapolate their local coefficients with the estimation kernel centered on each withheld observation, sharpened by an exponent chosen for each model and season by an internal validation within the training sample (\ref{B_apps}). Covariates of the kernel-based models are selected by AICc on the training sample, with and without an imposed time trend; the specification with the lower AICc is retained, the imposed trend being kept when the two differ by less than 10 points (\ref{B_apps}). Bold values indicate, per year and metric, the smallest error; $\dagger$ marks the models whose difference with the smallest error is not distinguishable from zero at the $5\%$ level, the $95\%$ percentile interval of the difference being obtained by a paired block bootstrap of the test sample ($2{,}000$ replicates, the fields as blocks).
\end{table}

\subsubsection{Results for Vaucluse house prices}\label{sec:vaucluse_results}

\paragraph{Specification}
To compare linear, locally linear and nonlinear specifications fairly, squared and cubic terms of the surface variables are candidates in all parametric models, and each model selects its covariates by a forward--backward stepwise search on an information criterion: the AIC for OLS and STVC, the AICc of a single-bandwidth GWR for the kernel-based models. OLS starts from all candidates except the time index and the month. The GWR search starts from the thirteen covariates that survive the collinearity screening, re-optimizes the bandwidth for each candidate set and retains eleven covariates (\ref{B_apps}). For GTWR and \texttt{tds\_mgtwr}, running the same search on their own criterion is prohibitive at this sample size: each candidate set would call for a two-dimensional bandwidth search, and the global AICc of the multiscale model, which rests on $n\times n$ projection matrices, is not computed for $n>40{,}000$ (\ref{app:timing}). Both therefore use the set selected by GWR, so that the kernel-based models differ only in their weighting schemes. STVC applies the same search to the AIC returned by \texttt{spmoran}, starting from this set, its temporal eigenvectors being built from the month index and the year of sale, and ends with twelve covariates, thirteen with the year of sale (Table~\ref{tab:vaucluse}). All locally weighted models use a cap of $3{,}000$ nearest neighbors out of the $40{,}598$ training sales. Preliminary sensitivity analyses showed diminishing returns in predictive accuracy as the cap increases, and this value was retained as a trade-off between accuracy and computational burden. The polynomial enrichment is not enough for a global model: OLS has the weakest accuracy (Table~\ref{tab:vaucluse}).

\paragraph{Comparison of estimators}
MARS and XGBoost, which receive the coordinates and time as inputs, improve markedly on OLS. The basis functions of MARS suggest that its gain comes mainly from smooth nonlinear effects of the year of sale and of the surface variables, rather than from interactions between space and time. Covariate-specific scales improve on the single space--time scale of GTWR by $2\%$ in RMSE and in MAE, and MGTWR has the lowest RMSE and MAE of all models, slightly ahead of XGBoost (RMSE $0.3540$ against $0.3560$, a mean over $20$ seeds, with a tree depth chosen without the 2022 sales). The paired block bootstrap of Table~\ref{tab:vaucluse} does not distinguish this RMSE margin from zero, nor the MAE margins over MARS and XGBoost, whereas every other difference with MGTWR is distinguishable. Unlike in the beet application, STVC is clearly behind MGTWR and GTWR (RMSE $0.3990$ against $0.3540$ and $0.3620$), and closer to OLS than to them. A plausible reason is that housing markets show sharp, localized discontinuities, for instance across neighborhood boundaries, rather than smooth global trends. The eigenvector basis of STVC and its regularization favor smoother patterns, whereas local kernels follow abrupt fine-scale heterogeneity without a superposition of high-frequency basis functions. Beyond accuracy, MGTWR provides a decomposition of spatial and temporal scales across covariates that purely predictive models do not deliver (Section~\ref{sec:vaucluse_scales}). These results can be read as conservative for MGTWR. Its covariates were selected by the AICc of a single-bandwidth GWR, a criterion that ignores covariate-specific scales, so that a covariate whose effect appears only at its own scale may have been discarded, and a selection adapted to MGTWR, as in the beet application, could be expected to improve its accuracy.

\begin{table}[htbp]
\centering
\caption{Out-of-sample performance (Vaucluse House Prices). XGBoost: mean over $20$ seeds, tree depth chosen within the training sample (\ref{B_apps}).}
\label{tab:vaucluse}

\begin{minipage}{0.85\linewidth}
\centering
\begin{tabular}{lccc}
\toprule
Model & RMSE & MAE & Selected Covariates\\
\midrule
OLS & 0.4183 & 0.3143 & 16\\
MARS u v t (R \texttt{earth}) & 0.3616 & 0.2466$^{\dagger}$ & 9\\
XGBOOST u v t (R \texttt{xgboost}) & 0.3560$^{\dagger}$ & 0.2483$^{\dagger}$ & 22 \\
\midrule
STVC (R \texttt{spmoran}) & 0.3990 & 0.3118 & 13\\
GTWR (R \texttt{mgwrsar}) & 0.3620 & 0.2500 & 11 \\
MGTWR (tds\_mgtwr, R \texttt{mgwrsar}) & \textbf{0.3540} & \textbf{0.2445} & 11\\
\bottomrule
\end{tabular}

\vspace{0.4em}
\footnotesize
\raggedright
\emph{Notes}: STVC covariates are selected by a forward--backward stepwise procedure on the AIC returned by \texttt{spmoran}, starting from the covariates of the kernel-based models; its temporal eigenvectors are built from the month index and the year of sale. The kernel-based models extrapolate their local coefficients with the estimation kernel centered on each test sale, sharpened by an exponent chosen by an internal validation on the 2021 sales (\ref{B_apps}). The count of XGBoost is its number of inputs: the eighteen candidates, the quarter entering as three indicators, and the two coordinates. Bold values indicate, per metric, the smallest error; $\dagger$ marks the models whose difference with the smallest error is not distinguishable from zero at the $5\%$ level, the $95\%$ percentile interval of the difference being obtained by a paired block bootstrap of the test sample ($2{,}000$ replicates, $2$~km squares as blocks). See \ref{B_apps} for estimation details.
\end{minipage}
\end{table}

\subsubsection{Spatial and temporal scales selected for Vaucluse house prices}
\label{sec:vaucluse_scales}

Beyond predictive accuracy, the bandwidths selected by \texttt{tds\_mgtwr} carry geographical information: each one tells over which neighborhood and which time span the effect of a covariate is smoothed. Since Gaussian weights decrease continuously, a bandwidth should be read as a smoothing scale rather than as the extent of an area where the effect is constant. Table~\ref{tab:vaucluse_bw} reports these scales for the MGTWR model of Table~\ref{tab:vaucluse}. Spatial bandwidths, defined as numbers of nearest sales so that local samples remain comparable between dense and sparse areas, are converted into distances, and spatio-temporal components are smoothed within a neighborhood of about $\pm 3$ years and $17$~km around each sale.

\begin{table}[htbp]
\centering
\caption{Bandwidths selected by \texttt{tds\_mgtwr} for Vaucluse house prices (model of Table~\ref{tab:vaucluse}, $n=40{,}598$). Spatial bandwidth: number $k$ of nearest sales and corresponding distance, median [interquartile range] over training sales; $^{\dagger}$spatio-temporal components, whose $k$ sales are counted within a space--time neighborhood of $3{,}000$ sales (about $\pm 3$ years). Temporal bandwidth: standard deviation of the Gaussian temporal kernel, in months. --: boundary choice, no weighting along that axis. Sign: dominant sign of the local marginal effect and share of training sales with that sign; for living area and land area, the marginal effect combines the linear, squared and cubic terms at the area of each sale. No significance is reported (Section~\ref{sec:pred_ac}).}
\label{tab:vaucluse_bw}
\footnotesize
\resizebox{\textwidth}{!}{%
\begin{tabular}{llrlrll}
\toprule
Term & Covariate & $k$ & Distance (km) & Months & Structure & Sign \\
\midrule
Intercept &  & 25 & 0.23 [0.16--0.39] & -- & spatial & -- \\
\texttt{anneemut} & year of sale & -- & -- & 37 & temporal & $-$ (78\%) \\
\texttt{sbati} & living area & 58 & 0.39 [0.28--0.70] & -- & spatial & $+$ (99\%) \\
\texttt{sbati\_2} & living area$^2$ & 555 & 4.74 [2.81--6.33]$^{\dagger}$ & 2 & spatio-temporal &  \\
\texttt{sbati\_3} & living area$^3$ & -- & -- & 90 & temporal &  \\
\texttt{sterr} & land area & 1{,}290 & 8.21 [6.43--11.48]$^{\dagger}$ & 90 & spatio-temporal & $+$ (94\%) \\
\texttt{sterr\_2} & land area$^2$ & 77 & 0.46 [0.33--0.83] & -- & spatial &  \\
\texttt{sterr\_3} & land area$^3$ & 2{,}264 & 5.97 [4.29--10.00] & -- & spatial &  \\
\texttt{ffnbppisci} & swimming pools & 974 & 3.80 [1.88--5.69] & -- & spatial & $+$ (100\%) \\
\texttt{dist\_ecole\_primaire} & distance to primary school & 419 & 4.01 [2.16--5.48]$^{\dagger}$ & 2 & spatio-temporal & $-$ (80\%) \\
\texttt{dist\_hopital} & distance to hospital & 2{,}264 & 12.49 [9.93--16.77]$^{\dagger}$ & 75 & spatio-temporal & $-$ (78\%) \\
\texttt{ffsdep} & outbuilding area & 2{,}264 & 5.97 [4.29--10.00] & -- & spatial & $+$ (100\%) \\
\bottomrule
\end{tabular}}
\end{table}

\paragraph{A variety of spatial and temporal scales}
The selected scales cover the whole range from the street block to the department, and no term is global: each one varies at least in space or in time. The finest spatial scales, a few hundred meters, are those of the price level, of the linear term of living area and of the squared term of land area, which shows that location premia differ even between adjacent neighborhoods. Swimming pools, proximity to schools and the squared term of living area vary over $4$ to $5$~km, roughly the size of a municipality, while outbuildings and land area vary over $6$ to $8$~km and proximity to hospitals over about $12$~km, the scale of groups of municipalities. In time, half of the components remain constant over the fifteen years, and the others drift over two months, about three years, six years or seven to eight years. The year of sale is the clearest case: it varies in time only, over about three years. Together with the purely spatial intercept, it splits the price level into a fine-grained location premium and a slowly evolving department-wide market trend.

\paragraph{Separating the linear and nonlinear parts of an effect}
Living area and land area enter through linear, squared and cubic terms, each with its own pair of scales. Because the powers of a size variable are strongly correlated, the scale of a single term depends on the others and is not interpreted on its own. Read together, however, the three terms show where and when the effect of size varies. For living area, the level of the marginal value of a square meter differs between nearby places, whereas its decreasing return with size, from a median effect of $0.65\%$ at $80$~m$^2$ to $0.49\%$ at $100$~m$^2$ and $0.28\%$ at $150$~m$^2$, changes over time as well as between places. This division of the effect between space and time was not imposed by the specification, since every term was offered both spatial and temporal candidates.
\subsubsection{Effect of living area on Vaucluse house prices}\label{sec:vaucluse_living_area}
\begin{figure}[htbp]
\begin{center}
\includegraphics[width=7in]{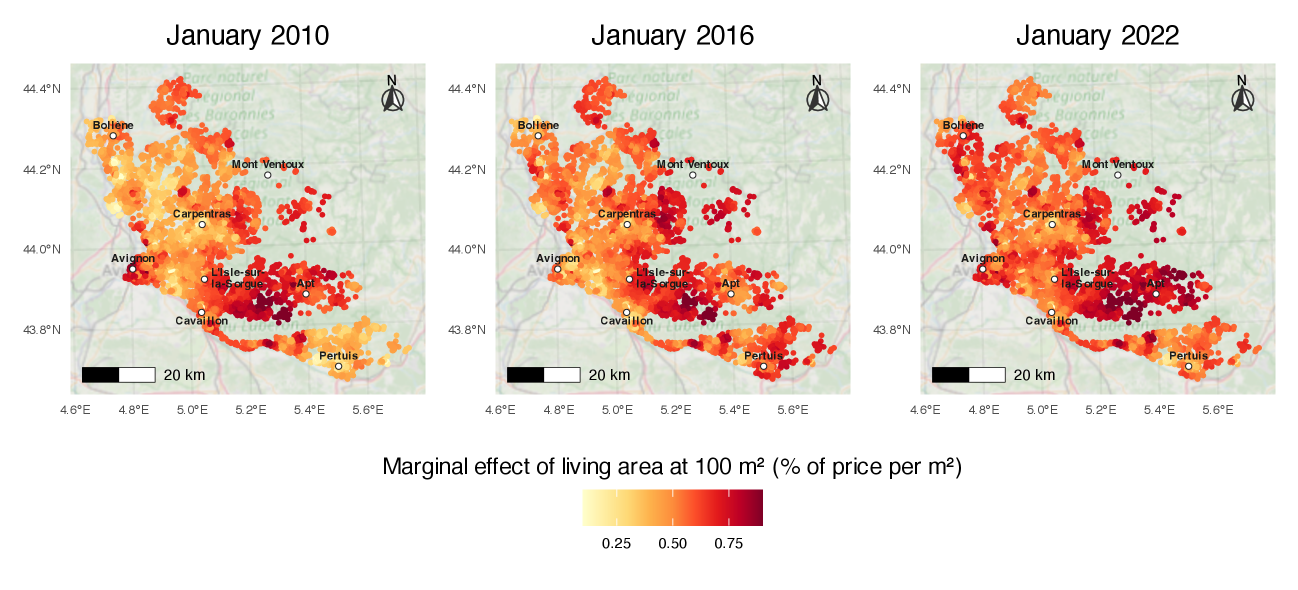}
\end{center}
\caption{Local marginal effect of living area estimated by MGTWR: percentage change in predicted price for one more square meter at $100$~m$^2$, other covariates fixed, at the sale locations in January 2010, 2016 and 2022 (MGTWR model of Table~\ref{tab:vaucluse}, fitted to the 2007--2021 sales: the January 2022 surface is a model-based extrapolation beyond the fitted period). Common color scale; values outside the 1st--99th percentiles of the three dates are shown at the scale bounds. Labeled points: places cited in Section~\ref{sec:vaucluse_living_area}. Basemap \copyright~OpenStreetMap contributors. Map lines delineate study areas and do not necessarily depict accepted national boundaries.}
\label{sbati}
\end{figure}

Figure~\ref{sbati} maps, at each sale location and at three dates, the percentage change in predicted price when living area increases from $100$ to $101$~m$^2$, other covariates being fixed. It is computed from the coefficients of the linear, squared and cubic terms of living area extrapolated by the MGTWR model of Table~\ref{tab:vaucluse}: with $b_1,b_2,b_3$ these coefficients on the original scale and the model being fitted to the logarithm of the price, the change in predicted log price is $\Delta\ell=b_1+b_2(101^2-100^2)+b_3(101^3-100^3)$ and the price change is $100\{\exp(\Delta\ell)-1\}\%$. The effect is positive at almost every sale location (all but $0.1$ to $0.2\%$ at each date). Its median rises from $0.48\%$ in January 2010 to $0.50\%$ in January 2016 and $0.55\%$ in January 2022, and its spatial dispersion is largest in 2010. The location of the strongest effects is partly persistent. Taking the 90th percentile of the three dates pooled ($0.71\%$) as a common threshold, the share of sales above it is $8.1\%$ in 2010, $7.7\%$ in 2016 and $14.2\%$ in 2022. At every date, high values are found around Apt, but with a varying weight: $27\%$ of local sales are above the threshold in 2010, $15\%$ in 2016 and $71\%$ in 2022. In 2010, a second cluster appears around Avignon ($12\%$). In 2016, high values spread to the foothills of Mont Ventoux east of Carpentras ($18\%$), to the area around Pertuis and to that between Cavaillon and L'Isle-sur-la-Sorgue ($10$ and $9\%$), while they almost vanish around Avignon ($1\%$). In 2022, they form a broad contiguous area around Apt, with secondary levels around Avignon ($15\%$) and Bollène in the north-west ($13\%$). At the end of the period, the marginal value of floor space thus concentrates over a wide inland area around Apt, between L'Isle-sur-la-Sorgue and Pertuis. Within each date, the finest local contrasts are consistent with the spatial scale of the linear living-area term, a few hundred meters, and the broader patterns with the scale of about $5$~km of the squared term (Table~\ref{tab:vaucluse_bw}). The changes between dates come from the time-varying terms of the polynomial.

\section{Conclusion}\label{sec:conclusion}
We have extended Top-Down Scale calibration to multiscale geographically and temporally weighted regression. The resulting estimator, \texttt{tds\_mgtwr}, selects a spatial and a temporal bandwidth for each covariate by a restricted search over ordered geometric grids, lets each component become spatial-only, temporal-only or global, and predicts out of sample by extrapolating the coefficients with sharpened kernels. As in MGWR backfitting, the in-sample error need not fall at every sweep, and the last state is returned, refitted at frozen bandwidths. Its support is therefore the evidence reported here rather than an optimality guarantee.

The simulations show that \texttt{tds\_mgtwr} recovers coefficient surfaces better than any competing estimator. It has the lowest mean coefficient error in every configuration of both designs, with a lead over STVC, the runner-up, that grows with sample size and signal strength and reaches $26\%$ in the cyclic design and $47\%$ in the non-cyclic one. Its predictions follow: from $n=1{,}000$ on it is the most accurate predictor, far ahead of MARS and XGBoost. It also outperforms the released MGTWR estimators, with a lower mean coefficient error than the Python package \texttt{mgtwr} and than the estimators of \citet{yu2024calibration} and MGTWR\_2SCALL \citep{wu2026efficient}, while being the fastest of the backfitting estimators in every timed configuration. Finally, its selected bandwidths lie close to the oracle scales of each coefficient, which no single-scale estimator achieves. Two limits remain: STVC stays more accurate on the curved-wave surface, where the oracle diagnostics point to the limits of isotropic smoothing, and with few observations, MGWR on the purely spatial coefficients and STVC remain competitive.

The applications show that letting coefficients vary in time as well as in space improves predictions only when the data document temporal variation well enough. For beet yellows severity, the best temporal specification varies with the season: covariate-specific scales (MGTWR) are the most accurate in 2020 and indistinguishable from the best model in 2021 and 2022, whereas a single space--time scale (GTWR) is the most accurate in 2023 and level with XGBoost in 2022. A complete application of MGTWR to spatio-temporal model averaging of beet yellows forecasts, in which local coefficients serve as aggregation weights that adapt jointly to spatial context and temporal phase, is developed in a separate study under submission \citep{martinez2026}. For Vaucluse house prices, MGTWR has the lowest RMSE and MAE, slightly ahead of XGBoost, the best of the machine-learning benchmarks, by a margin within bootstrap uncertainty, and its selected scales carry geographical information: a fine-grained location premium, a department-wide market trend, and, for living area, a level that differs between nearby places and a decreasing return that also changes over time. These scales describe smoothing conditional on observation density and model selection, and require geographical interpretation alongside the sampling pattern. On the beet yellows severity data, exact standard errors and their local approximation differ by up to a factor of five, which motivates caution in significance mapping.

For the practitioner, \texttt{tds\_mgtwr} changes what can be done with MGTWR. The released estimators either accepted only balanced panels or became very slow as the sample or the number of covariates grew, and none of them could predict. \texttt{tds\_mgtwr} handles irregular space--time point patterns, repeated locations and linear or cyclic time, and it was fitted here on $40{,}598$ sales with eleven covariates in under an hour on ten cores, with new observations predicted in seconds. The analyst no longer has to decide in advance which effects vary in space, in time or in both, or to accept a single scale for all of them: the boundary candidates let the data decide, covariate by covariate, and the selected scales can be read as a description of the process. The analyst also gains a better control of degenerate solutions and concurvity than a criterion-driven search alone can give: the pre-selection and the grid floors rule out scales at which a covariate hardly varies locally, saturated candidates are discarded, local systems are solved by a rank-revealing method, and a concurvity check between fitted components guides covariate selection (\ref{B_apps}). The calibration keeps the familiar information-criterion logic of the GWR family, examines only a few dozen bandwidth pairs at each update, and is available in the R package \texttt{mgwrsar}, which brings multiscale spatio-temporal estimation to samples of tens of thousands of observations on standard hardware.

The approach has several limitations. The most important is that the hat-matrix quantities remain costly for large samples. At the size of the house-price application, neither the exact standard errors nor the global AICc are tractable, so that covariates cannot be selected by a criterion adapted to MGTWR (Section~\ref{sec:vaucluse_results}) and the univariate approximation is no substitute for inference (\ref{B_inf}). The chunk-wise computation of \citet{li2020computational} proposed for MGWR does not lift this limit. By computing the columns of the $n \times n$ projection matrices block by block, it reduces the memory needed at any one time and leaves the result unchanged, but all the columns are still computed, with up to $O(n^3)$ operations at each update of a component, and for a given number of blocks the storage remains of order $n^2$. Efficient approximations of the effective number of parameters and of the standard errors therefore remain to be developed. Second, the separable kernel favors clarity and computational efficiency but excludes explicit space--time interaction kernels, which may matter for propagation-type dynamics or localized temporal shocks. Third, bandwidths are global per covariate and kernels are isotropic, whereas coefficient surfaces may call for scales or orientations that vary across space or season. Fourth, bandwidths take values on discrete grids, whose rounding to whole time units makes them coarser at short temporal bandwidths (\ref{B_controls}). Fifth, the robustness of the calibration to strong local collinearity has been tested only on the simulated design of \ref{C_order}. 


The main extension is a scalable approximation of the hat-matrix quantities, which would bring valid inference and a covariate selection adapted to MGTWR to large samples. Covariate selection based on cross-validation, which needs no hat matrix, could also be designed and evaluated as an alternative at these sample sizes. Other extensions include spatio-temporally adaptive bandwidths, for which the adaptive version of \texttt{tds\_mgwr} (\emph{atds\_mgwr}; \citealt{geniaux2026top}) provides a natural foundation, and anisotropic kernels, which the oracle diagnostic of \ref{B_mc} suggests for wave-like surfaces. A continuous refinement of the bandwidths between grid levels and multistart strategies, to assess the sensitivity to the update order and to local collinearity, would also be useful.

\section{Code availability}\label{sec:code}
The \texttt{tds\_mgwr} and \texttt{tds\_mgtwr} estimators are implemented in the R package \texttt{mgwrsar} (version 1.4.4, CRAN). The scripts that reproduce the Monte Carlo experiments of Section~\ref{sec:simuls} and \ref{B_mc} and the Vaucluse house-price application of Section~\ref{sec:realdata}, together with the per-replication results of the published runs and the R port of the local estimator of \citet{yu2024calibration} with its validation test and of their global search, are provided to the reviewers as supplementary material and will be made public upon publication at \url{https://github.com/ggeniaux/tds_mgtwr_replication}, archived on Zenodo. They run without a computing cluster and include a test mode on a reduced number of replications. The beet yellows severity application is not covered, its data being confidential.

\section{Declaration of generative AI and AI-assisted technologies in the manuscript preparation process}

During the preparation of this manuscript, the authors used generative AI tools (Claude and ChatGPT) to assist with English language editing and to support the development of code within the \texttt{mgwrsar} R package. All outputs generated with the assistance of these tools were reviewed, edited, and validated by the authors. The authors take full responsibility for the content of the manuscript and for any decisions made in its preparation.

\section{Declaration of competing interest}
The authors declare that they have no known competing financial interests or personal relationships that could have appeared to influence the work reported in this paper.

\section{Acknowledgements} Funding: This work was supported by the French National Research Agency [grant number ANR-20-PCPA-0002] (BEYOND project, Building epidemiological surveillance and prophylaxis with observations both near and distant) and 	(SEPIM project, Surveillance, Évaluation, Prévision, Interpolation et Mitigation des risques relatifs à la jaunisse de la betterave). The funding sources had no involvement in the study design, in the collection, analysis and interpretation of data, in the writing of the report, or in the decision to submit the article for publication. Numerical computations were performed using the INRAE biosp-cluster  (https://biosp-cluster.mathnum.inrae.fr/ )  and the INRAE MIGALE bioinformatics facility  (MIGALE, INRAE, 2020. Migale bioinformatics Facility, doi: \url{10.15454/1.5572390655343293E12}).

\bibliographystyle{elsarticle-harv}
\bibliography{tds_mgtwr}

\appendix
\counterwithin*{table}{section}
\counterwithin*{figure}{section}

\section{Per-covariate coefficient extrapolation}
\label{A_extrapolation}
After estimation, MGTWR delivers, for each covariate $k=1,\ldots,K$ (with $k=1$ the intercept), local coefficients $\widehat{\beta}_k(j)$ at each training observation $j\in S$ and a selected bandwidth pair $(h_{k,s}^{\star},h_{k,t}^{\star})$, together with the temporal kernel as configured at estimation (symmetric or unilateral, linear or cyclic). For a prediction point $o\in O$ at time $t_o$, with covariate vector $\mathbf{x}_o=(x_{o1},\ldots,x_{oK})^\top$, the per-covariate pre-weights use the same separable kernel as the estimation step, $d_{jo}$ being the spatial distance between $j$ and $o$:
\[
\widetilde{K}_k(o,j)=K^{(s)}\!\left(d_{jo};\,h_{k,s}^{\star}\right)\times K^{(t)}\!\left(t_o,t_j;\,h_{k,t}^{\star}\right),\qquad
\widetilde{W}_k(o,j)=\frac{\widetilde{K}_k(o,j)}{\sum_{j'\in S}\widetilde{K}_k(o,j')}.
\]
We then \emph{sharpen} the kernel by a power transform and renormalization:
\[
\breve{W}_k(o,j)=\big[\widetilde{W}_k(o,j)\big]^{\gamma},\qquad
W_k(o,j)=\frac{\breve{W}_k(o,j)}{\sum_{j'\in S}\breve{W}_k(o,j')},
\]
where the exponent $\gamma\geq1$ is a tuning parameter: $\gamma=8$ in the Monte Carlo predictions (Section~\ref{sec:simuls}), and a value chosen for each estimator and season by an internal validation within the training sample in the applications (\ref{B_apps}). The pre-weight matrices are never formed as dense arrays: for each covariate in turn, the weights are stored as a sparse matrix with at most $NN$ non-zero entries per prediction point. The extrapolated coefficient and the prediction are
\[
\widehat{\beta}_k(o,t_o)=\sum_{j\in S} W_k(o,j)\,\widehat{\beta}_k(j),\qquad
\widehat{y}_o=\sum_{k=1}^K x_{ok}\,\widehat{\beta}_k(o,t_o).
\]
This rule reuses the learned scales of each covariate and \emph{does not} re-optimize any prediction bandwidth. MGWR is the special case without a temporal axis, $K^{(t)}\equiv1$, in which the pre-weights reduce to the spatial kernel at $h_{k,s}^{\star}$. The same reduction applies component by component at the boundary choices of Section~\ref{sec:algo}: only the spatial marginal remains at the temporal sentinel, only the temporal marginal at the spatial sentinel, and a global component reuses its coefficient for all $o$.

\section{Order Effects and Convergence Properties in Multiscale Backfitting}
\label{C_order}
The multiscale backfitting of MGWR and of its top-down scale (TDS) extension is related to a block coordinate descent on a non-convex information criterion (AICc) over a discrete grid of candidate bandwidths; the analogy motivates the discussion below but, as stated in Section~\ref{sec:algo}, no convergence result is claimed for the algorithm itself. In the classical additive model, backfitting is a Gauss--Seidel iteration over fixed linear smoothers: with a quadratic objective and a unique minimizer, the order of updates affects the rate of convergence but not the solution \citep{buja1989linear}. Neither condition holds in MGWR and \texttt{tds\_mgtwr}: the smoothers depend on the bandwidths, the AICc is non-convex in the bandwidth vector, and the grid is discrete. For non-convex problems, cyclic block coordinate descent converges only to a stationary point, which may depend on the update order and on the initialization \citep{tseng2001convergence}. The allocation of large-scale structure across covariates is therefore path-dependent: early-updated covariates absorb variance components that shape the partial residuals of the later ones, particularly under local collinearity, and in \texttt{tds\_mgtwr}, where each block is a pair of spatial and temporal bandwidths, the search space is larger still.

Two alternatives to a fixed cyclic order address this path dependence. Randomized orders improve convergence rates in expectation in convex settings \citep{nesterov2012efficiency} and, for non-convex block optimization, still converge only to stationary points but can mitigate systematic ordering effects \citep{razaviyayn2013unified}. Gauss--Southwell-type rules update first the coordinates with the largest marginal impact on the objective \citep{nutini2015coordinate}; their acceleration results, established for smooth and strongly convex problems, do not carry over to a discrete non-convex AICc, but they motivate a heuristic. The importance-driven order of Section~\ref{sec:algo} is such a deterministic analogue.

Table~\ref{tab:order_collin} compares the three orderings on $100$ replications of the central configuration of Section~\ref{sec:simuls} ($n=1{,}000$, $\rho=0.9$, cyclic design), with the covariates $X_1$, $X_2$, $X_3$ independent, as in the Monte Carlo experiments, or made equicorrelated at $0.5$ and $0.8$ by a Cholesky transformation of the same draws, the coefficient surfaces being unchanged and the noise recalibrated to the same signal index. Correlation among covariates degrades every estimator alike (the mean coefficient RMSE rises from $0.77$ to $0.91$), but the ordering does not matter on average: the three orders agree within $0.1\%$ in mean coefficient RMSE and prediction RMSE at every correlation level, below one Monte Carlo standard error ($0.004$ to $0.005$). Path dependence is nevertheless visible replication by replication: the cyclic and random orders end at other bandwidths than the importance-driven order in $4$ and $11\%$ of replications with independent covariates, and in $24$ and $20\%$ at correlation $0.8$, where the coefficient RMSE of a single replication can move by up to $2\%$ (cyclic) and $8\%$ (random). The random order needs about one sweep more. All reported results, simulations and applications, use the importance-driven default.

\begin{table}[htbp]
\centering
\caption{Update order and correlated covariates: \texttt{tds\_mgtwr} on $100$ replications of the central configuration ($n=1{,}000$, $\rho=0.9$, cyclic design), covariates independent ($\rho_x=0$) or equicorrelated at $\rho_x=0.5$ and $0.8$. Mean RMSE of each coefficient and their mean, out-of-sample prediction RMSE, mean number of sweeps, and share of replications in which the selected bandwidths differ from those of the importance-driven order.}
\label{tab:order_collin}
\footnotesize
\resizebox{\textwidth}{!}{%
\begin{tabular}{llcccccccc}
\toprule
Update order & $\rho_x$ & $\beta_1$ & $\beta_2$ & $\beta_3$ & $\beta_4$ & Mean & Pred. & Sweeps & Bandwidths $\neq$ (\%) \\
\midrule
Importance (default) & 0.0 & 0.701 & 0.703 & 0.616 & 1.075 & 0.774 & 2.611 & 26.4 & -- \\
Cyclic & 0.0 & 0.701 & 0.703 & 0.616 & 1.075 & 0.774 & 2.611 & 26.2 & 4 \\
Random & 0.0 & 0.701 & 0.703 & 0.616 & 1.074 & 0.774 & 2.610 & 27.5 & 11 \\
\midrule
Importance (default) & 0.5 & 0.709 & 0.729 & 0.683 & 1.148 & 0.817 & 2.681 & 27.1 & -- \\
Cyclic & 0.5 & 0.709 & 0.729 & 0.683 & 1.148 & 0.817 & 2.681 & 26.8 & 2 \\
Random & 0.5 & 0.709 & 0.729 & 0.683 & 1.147 & 0.817 & 2.680 & 28.7 & 12 \\
\midrule
Importance (default) & 0.8 & 0.709 & 0.779 & 0.850 & 1.316 & 0.913 & 2.694 & 28.3 & -- \\
Cyclic & 0.8 & 0.708 & 0.780 & 0.849 & 1.316 & 0.913 & 2.694 & 28.0 & 24 \\
Random & 0.8 & 0.708 & 0.779 & 0.851 & 1.315 & 0.913 & 2.694 & 29.4 & 20 \\
\bottomrule
\end{tabular}}
\end{table}

\section{Numerical controls of the TDS calibration}
\label{B_controls}

\paragraph{Numerical controls of the TDS calibration}
Table~\ref{tab:tds_controls} lists the grid bounds, stopping rules and safeguards of \texttt{tds\_mgtwr} (Section~\ref{sec:algo}), with the values used in all reported experiments, which are the defaults of \texttt{mgwrsar} 1.4.4 except for $\varepsilon$ (default $10^{-3}$).

\begin{table}[htbp]
\centering
\caption{Numerical controls of \texttt{tds\_mgtwr} used in all experiments (defaults of \texttt{mgwrsar} 1.4.4, except for $\varepsilon$). RMSE: in-sample root mean squared error of the complete fit after a sweep.}
\label{tab:tds_controls}
\small
\begin{tabular}{p{4.6cm} p{10.4cm}}
\toprule
Control & Value \\
\midrule
Grid length $M$ & $20$ (simulations), $25$ (applications); geometric sequences on both axes \\
Spatial bounds & upper: whole sample, or the neighbor cap when one is set (largest pairwise distance for non-adaptive kernels); lower: at least $K+1$ points for compact-support kernels, a small fixed floor for Gaussian kernels \\
Temporal bounds & upper: largest temporal gap within the neighborhoods considered, $C/2$ for cyclic kernels ($182$ days for $C=365$); lower: the $2\%$ quantile of the positive pairwise gaps; values rounded to whole time units (days or periods), so that consecutive short bandwidths may differ by one unit \\
Candidate pre-selection & optional: local weighted variance above $5\%$ of the global variance at $90\%$ of the locations; used in the applications, not in the simulations \\
Search freeze & when no bandwidth pair has changed for four consecutive sweeps, the candidate search is skipped and the components are refitted at their bandwidths \\
Saturation guard & candidate excluded when its local fit absorbs more than $86\%$ of the residual variance of the current complete fit (residual variance below $\exp(-2)$ times that of the complete fit); if all are excluded, smallest AICc kept with a warning; counts returned with the fit \\
Convergence & relative change of the RMSE between successive sweeps below $\varepsilon=10^{-4}$ for three consecutive sweeps \\
Patience & stop after five consecutive sweeps without improvement of the best RMSE \\
Oscillation & stop when the sign of the RMSE change alternates over at least three consecutive sweeps with a mean absolute relative change below $\varepsilon$ over the last two \\
Ping-pong & stop when the same two bandwidth configurations alternate three times \\
Iteration limit & $100$ sweeps \\
Polishing & bandwidths frozen, at most $3$ further sweeps, stopped when the relative RMSE change is below $\varepsilon$; the terminal state is returned (the state of smallest RMSE over the sweeps, returned by versions up to 1.4.2, remains available as an option) \\
Update order & importance-driven (standard deviation of the fitted contribution, intercept first); cyclic and random orders available \\
\bottomrule
\end{tabular}
\end{table}

\begin{figure}[htbp]
\begin{center}
\includegraphics[width=5in]{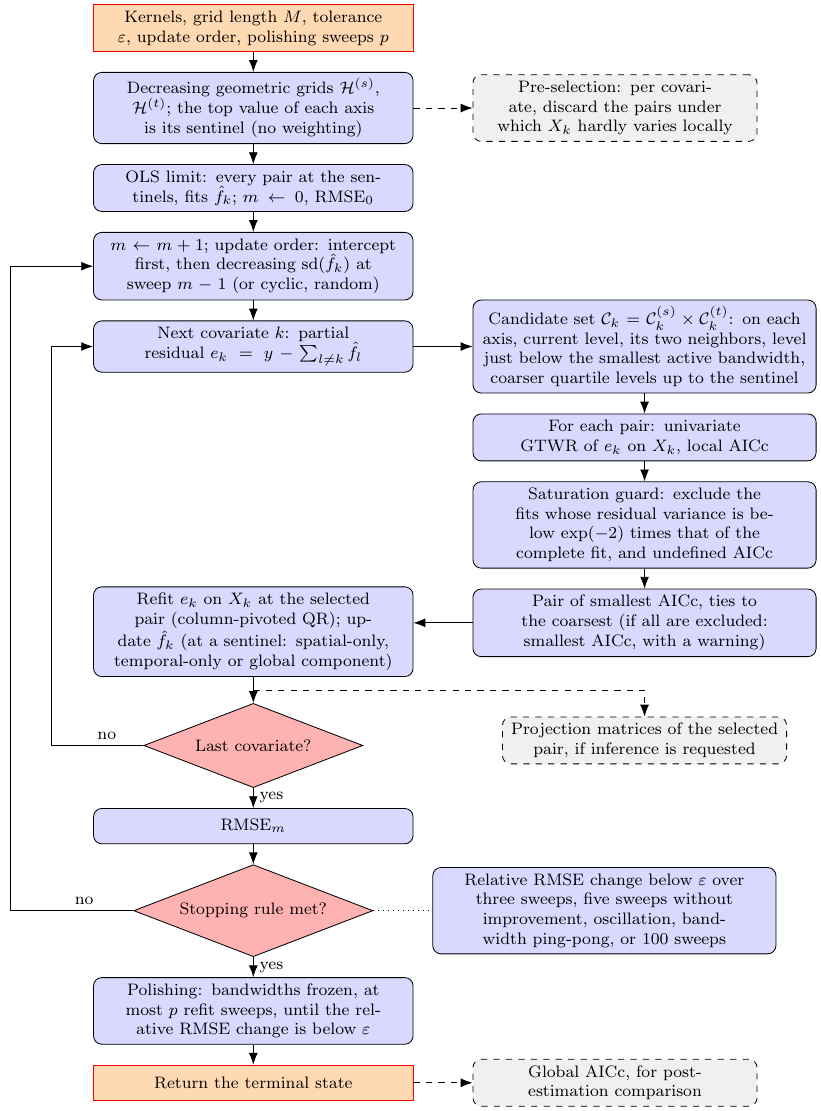}
\end{center}
\caption{Schematic overview of the \texttt{tds\_mgtwr} calibration, with the values of Table~\ref{tab:tds_controls} and the importance-driven update order (see also Algorithm~\ref{alg:tds_mgtwr}). Dashed boxes: optional steps (candidate pre-selection, projection matrices and global AICc when inference is requested).}\label{flow1}
\end{figure}

\paragraph{Sensitivity of the TDS calibration to grid granularity}
With the default bounds and numerical controls of Section~\ref{sec:algo} held fixed, the granularity $M$ determines the bandwidth grids. To assess robustness, we re-estimated \texttt{tds\_mgtwr} on $100$ replications of the DGP of Section~\ref{sec:mc_design} ($n=1{,}000$, $\rho=0.9$, cyclic design, Gaussian spatial and periodic temporal kernels) for $M\in\{10,20,30\}$. Table~\ref{tab:grid_sensitivity} reports, for each covariate, the median selected bandwidth pair and the mean coefficient RMSE with its Monte Carlo standard error. Three observations stand out. First, the coarsest grid costs accuracy and the finest adds none: relative to $M=20$, $M=10$ raises the mean RMSE by $0.7$ to $1.7\%$ for $\beta_1$, $\beta_2$ and $\beta_3$ and by $6\%$ for $\beta_4$ (paired differences, each at least three standard errors from zero), whereas $M=30$ changes it by less than $0.3\%$, within one standard error, at the cost of $30\%$ more sweeps ($34$ against $26$); the in-sample RMSE follows the same pattern ($1.68$, $1.60$, $1.60$). Second, the selected bandwidths move only within adjacent grid steps ($32$--$37$ and $21$--$23$ days for the median temporal scales of $\beta_1$ and $\beta_2$; spatial distances stable within $10\%$ for all covariates). Third, the two purely spatial coefficients $\beta_3$ and $\beta_4$ select the \emph{maximal} temporal bandwidth in most replications ($77$ and $92\%$ at $M=20$, $76$ and $96\%$ at $M=30$, $81$ and $69\%$ at $M=10$): the boundary-solution mechanism of Section~\ref{sec:algo} collapses their temporal component, recovering the spatial-only nature of these coefficients regardless of grid granularity. In short, $M=20$, the lower end of the range recommended by \citet{geniaux2026top}, is enough on this design: a coarser grid loses accuracy, mainly on $\beta_4$, a finer one only adds sweeps, and the selected scales barely move. This range can thus serve as a rule of thumb for future MGTWR applications. The script is included in the replication repository.

\begin{table}[htbp]
\centering
\caption{Sensitivity of \texttt{tds\_mgtwr} to the grid granularity $M$ ($100$ replications of the DGP of Section~\ref{sec:mc_design}, $n=1{,}000$, $\rho=0.9$, cyclic design). $h^{(s)}$: median selected spatial bandwidth (distance, Gaussian kernel); $h^{(t)}$: median selected temporal bandwidth (days; 182 = maximal value of the periodic kernel, i.e., boundary solution collapsing to a spatial-only component); RMSE: mean coefficient RMSE as defined in Section~\ref{sec:mc_design}, Monte Carlo standard error in parentheses.}
\label{tab:grid_sensitivity}
\resizebox{\textwidth}{!}{%
\begin{tabular}{lccc c ccc c ccc}
\toprule
 & \multicolumn{3}{c}{$M=10$} & & \multicolumn{3}{c}{$M=20$} & & \multicolumn{3}{c}{$M=30$} \\
\cmidrule{2-4}\cmidrule{6-8}\cmidrule{10-12}
 & $h^{(s)}$ & $h^{(t)}$ & RMSE & & $h^{(s)}$ & $h^{(t)}$ & RMSE & & $h^{(s)}$ & $h^{(t)}$ & RMSE \\
\midrule
$\beta_1$ & 0.081 & 32 & 0.713 (0.006) & & 0.085 & 35 & 0.701 (0.005) & & 0.087 & 37 & 0.700 (0.005) \\
$\beta_2$ & 0.109 & 23 & 0.708 (0.005) & & 0.117 & 21 & 0.703 (0.005) & & 0.109 & 22 & 0.701 (0.005) \\
$\beta_3$ & 0.061 & 182 & 0.623 (0.005) & & 0.052 & 182 & 0.616 (0.005) & & 0.051 & 182 & 0.615 (0.005) \\
$\beta_4$ & 0.036 & 182 & 1.141 (0.006) & & 0.031 & 182 & 1.075 (0.007) & & 0.031 & 182 & 1.074 (0.007) \\
\midrule
in-sample RMSE & \multicolumn{3}{c}{1.683 (0.013)} & & \multicolumn{3}{c}{1.602 (0.012)} & & \multicolumn{3}{c}{1.603 (0.011)} \\
\bottomrule
\end{tabular}}
\end{table}

\section{Monte Carlo settings}
\label{B_mc}

\paragraph{Construction of $\beta_3$}
The coefficient $\beta_3$ of Section~\ref{sec:mc_design} mimics an accessibility gradient and stacks three layers:
\[
\beta_3(u_i,v_i)\;=\;\underbrace{2.1\,k_{0i}}_{\text{road}}\;+\;\underbrace{\sum_{j=1}^{5}a_j\,k_{ji}}_{\text{towns}}\;+\;\underbrace{1.5\,\varsigma_i\prod_{j=0}^{5}\big(1-k_{ji}\big)}_{\text{background}}.
\]
Road and towns share one profile, $k(d;r)=(1-d/r)_+^2$, which equals $1$ at the feature and falls to $0$ at distance $r$. The road is the curve $v=g(u)=0.2+0.6\,u+0.12\sin(3\pi u)$, crossing the study area from $(0,0.2)$ to $(1,0.8)$, and $k_{0i}=k(d_{0i};0.12)$, with $d_{0i}$ the distance to the road: a ridge of height $2.1$ and half-width $0.12$. The five towns lie on the road at $u_j\in\{0.12,0.30,0.50,0.70,0.88\}$, with radii $r_j\in\{0.06,0.08,0.58,0.07,0.66\}$ and heights $a_j\in\{4.2,5.6,7.0,3.5,4.9\}$, and $k_{ji}=k(d_{ji};r_j)$, with $d_{ji}$ the distance to the center of town $j$: three narrow peaks and two broad halos. The background takes the sign $\varsigma_i=-1$ north-west of the road and $+1$ south-east of it. Its product vanishes on the road and at the town centers and equals $1$ beyond the reach of all features, so that the level $\pm1.5$ appears only where road and towns have no influence and $\beta_3$ stays continuous across the road.

\paragraph{Noise level}
Over the first $100$ replications of each cell, the noise scale $a$ is $3.60$ to $3.66$ at $\rho=0.7$ and $1.82$ to $1.88$ at $\rho=0.9$, in both designs. The systematic component has mean $3.0$ and standard deviation $4.7$, hence centered signal-to-noise variance ratios of $1.6$ to $1.7$ and $6.2$ to $6.6$. The out-of-sample RMSE of the oracle equals $a$.

\paragraph{Estimators of the point-pattern design}
Table~\ref{tab:mc_settings} gives the settings, every estimator running on a single core. \texttt{tds\_mgtwr} and GTWR share the same separable, symmetric kernel, periodic with a $365$-day cycle in the cyclic design, so that they differ only in their bandwidths. The Python package is started from a nearly global fit, the counterpart of the OLS start of \texttt{tds\_mgtwr}, because its default start, a single-bandwidth GTWR search with all regressors, fails with a singular local matrix in most replications. It was not run at $n=5{,}000$: at $n=1{,}000$, one fit takes about $20$ times as long as with \texttt{tds\_mgtwr} (medians of $12$ and $15$ minutes for the two signal indices, against less than one minute). STVC runs through the modified implementation included in the replication repository, which differs from \texttt{spmoran::resf\_vc} by about $1$--$2\%$ in our implementation check. MARS and XGBoost are fixed rather than broadly tuned: at $n=600$, a five-fold training fold holds about $480$ observations, which \texttt{min\_child\_weight}$=200$ strongly constrains.

\begin{table}[htbp]
\centering
\caption{Settings of the point-pattern experiments (Section~\ref{sec:simuls}).}
\label{tab:mc_settings}
\footnotesize
\begin{tabular}{@{}>{\raggedright\arraybackslash}p{3.3cm} >{\raggedright\arraybackslash}p{6.2cm} >{\raggedright\arraybackslash}p{7.0cm}@{}}
\toprule
Estimator & Kernel or basis & Selection and tuning \\
\midrule
GWR & fixed Gaussian, spatial & bandwidth by AICc over $[0,d_{\max}]$ \\
GTWR & fixed Gaussian, space $\times$ time & pair by AICc, on a grid over $[d_{\min},d_{\max}]$ and $[4,182]$ days (cyclic) or $[10,1{,}460]$ days (non-cyclic) \\
\texttt{tds\_mgwr}, \texttt{tds\_mgtwr} & as GWR and GTWR, per covariate & $M=20$, OLS start, importance order, $\varepsilon=10^{-4}$ \\
\texttt{mgtwr} (Python, 2.0.5) & fixed Gaussian on a single space--time distance, time in years & bandwidth in $[0.02,1.5]$ (three decimals), $\tau$ in $[0,16]$ (six decimals); start at $\mathrm{bw}=1.5$, $\tau=10^{-6}$ \\
STVC & up to $500$ spatio-temporal eigenvectors with interactions; temporal coordinates: cosine and sine of the day of year (cyclic), scaled absolute day (non-cyclic) & eigenvector cleaning \texttt{clean\_meigen}, fitting function \texttt{resf\_vc\_correct} \\
MARS & hinge basis, interaction degree two; inputs: covariates, coordinates, day of year, year & \texttt{nfold}$=10$ \\
XGBoost & trees of depth four; same inputs as MARS & learning rate $0.1$, \texttt{min\_child\_weight}$=200$, rounds by five-fold cross-validation with early stopping \\
\bottomrule
\end{tabular}
\end{table}

\paragraph{Package version}
The TDS rows of Tables~\ref{tab::monte_carlo_results}, \ref{tab::panel_rmse} and \ref{tab::panel_bw} were computed with \texttt{mgwrsar} 1.4.1. Recomputing $30$ replications of the $n=1{,}000$, $\rho=0.7$ cell with version 1.4.4, in both designs, reproduces them to within $0.2\%$ in mean coefficient RMSE and $0.1\%$ in prediction RMSE, with identical bandwidths in $90$ to $100\%$ of the replications. On the balanced panel, $10$ replications are reproduced exactly (same bandwidths, coefficient RMSE within $10^{-4}$).

\paragraph{Oracle diagnostic of the purely spatial coefficients}
To locate the gap between \texttt{tds\_mgtwr} and STVC on the purely spatial coefficients (Section~\ref{sec::monte_carlo_results}), we replaced coefficients by their true values step by step, on $20$ replications of the central configuration ($n=1{,}000$, $\rho=0.9$, cyclic design). \texttt{tds\_mgtwr} reaches an RMSE of $0.62$ on $\beta_3$ and $1.07$ on $\beta_4$, against $0.67$ and $0.90$ for STVC. The oracle fits each coefficient on its exact partial residual, the other coefficients being set to their true values, with the isotropic Gaussian bandwidths that minimize the RMSE of $\hat\beta_k$. The oracle RMSE is $0.53$ on $\beta_3$ and $0.93$ on $\beta_4$. The gap between \texttt{tds\_mgtwr} and the oracle comes first from the errors on $\beta_1$ and $\beta_2$ carried over by backfitting ($0.04$ on $\beta_3$, $0.09$ on $\beta_4$), then from the joint estimation of $\beta_3$ and $\beta_4$ ($0.04$ each), while selecting the bandwidths by AICc rather than by the RMSE costs about $0.01$. These are descriptive contrasts on $20$ replications, not a unique additive decomposition.

On $\beta_3$, whose road and towns are local features of several sizes, the oracle lies well below STVC, and \texttt{tds\_mgtwr} already does. On $\beta_4$, a set of rings of short and constant wavelength covering the whole domain, even the oracle remains above STVC, whose eigenvectors represent such a regular global pattern parsimoniously. A kernel oriented along the direction of variation lowers the oracle RMSE of $\beta_4$ to $0.84$, below STVC, but a single orientation per coefficient gains only $10\%$ because that direction changes across space, and a local-linear fit gains $5\%$ ($0.89$). On $\beta_3$ neither helps ($2\%$ at most). We return to kernel shapes in the Conclusion.

\paragraph{Construction of the balanced panel}
The panel of Section~\ref{sec::comparator_mc} transposes the point-pattern design to the only sampling scheme that the released codes accept: $N=100$ units observed at $T=20$ equally spaced periods ($n=2{,}000$), with the model of Section~\ref{sec:mc_design} and the phase $s_t=(t-1)/(T-1)$. This requires three adaptations. First, the units sit on a regular $10\times10$ lattice over $[0,1]^2$, each node being jittered uniformly by at most $3\%$ of the lattice step, with a new draw per replication: on an exact lattice, many units are equidistant from a given one, so that the $k$-th nearest neighbor, which sets the radius of an adaptive kernel, would be ambiguous. Second, $\beta_1$ and $\beta_2$ are taken in linear time. Third, the purely spatial surfaces are adapted to the lattice step, about $0.11$, which is coarser than some of their features: the half-width of the road and the radii of the three small towns of $\beta_3$ are multiplied by $2.5$, so that each small town covers $6$ to $10$ units instead of one, and $\beta_4(u,v)=4\sin\!\big(6\sqrt2\,\lVert(u,v)-(1,0)\rVert\big)+2(u-0.5)$ has a wavelength of $0.74$ instead of $0.30$. The noise is scaled so that $\rho=\sum\eta_{it}^2/\sum y_{it}^2=0.9$, a variance ratio $\widehat{\mathrm{Var}}\{\eta\}/\sigma^2$ of about $5$ to $6$, and $1{,}000$ replications are drawn.

\paragraph{Estimators of the balanced panel}
Table~\ref{tab:panel_settings} gives the settings. Since all units are observed at each period, spatial bandwidths below the distance $d_1$ to the nearest distinct unit remain identifiable through each unit's own observations, so that the spatial grid of \texttt{tds\_mgtwr} is extended geometrically down to $0.5\,d_1$. The released code of \citet{yu2024calibration} contains two estimators. The local one is the backfitting of \citet{wu2019multiscale}, in which the bandwidths of each component are re-selected at each update, and whose failures in space--time motivate their global search (their Section~3). Its genetic search over the pair of each component, unseeded in the released driver, is replaced by an exhaustive enumeration of the same integer grid, which returns the best pair of its criterion at each update and makes the comparator deterministic, conditional on the data and the initialization.

\begin{table}[htbp]
\centering
\caption{Settings of the balanced-panel comparison (Section~\ref{sec::comparator_mc}).}
\label{tab:panel_settings}
\footnotesize
\begin{tabular}{@{}>{\raggedright\arraybackslash}p{3.5cm} >{\raggedright\arraybackslash}p{4.2cm} >{\raggedright\arraybackslash}p{5.9cm} >{\raggedright\arraybackslash}p{2.7cm}@{}}
\toprule
Estimator & Kernels & Bandwidth search & Code \\
\midrule
\texttt{tds\_mgtwr} & fixed bi-square on both axes, non-cyclic & default settings, spatial grid extended to $0.5\,d_1$ & \texttt{mgwrsar} \\
Yu--Fotheringham, local \texttt{mgtwr.m} & released bi-square: adaptive neighbor count in space, window in periods & univariate AICc of each partial residual, exhaustive enumeration of $[5,N]\times[1,T]$ at each update & R port of the released files \\
Yu--Fotheringham, global \texttt{mgtwr\-global.m} & as the local estimator & genetic algorithm on all $2K$ bandwidths, each candidate fitted by backfitting at fixed bandwidths and scored by the AICc of the complete model; authors' integer bounds and stopping rule, population $80$ for $K=4$, replication number as seed & R port with the \texttt{GA} package; first $100$ replications \\
MGTWR\_2SCALL, released driver & adaptive product bi-square, $c_s=c_t=0.8$, spatial floor lowered from $30$ to $5$ units & one golden section per bandwidth, the other being held fixed & released MATLAB code \\
MGTWR\_2SCALL, alternating search & as the released driver & spatial and temporal golden sections repeated until the pair stabilizes (Section~3 of \citet{wu2026efficient}) & our loop around the unmodified fitting routines \\
\bottomrule
\end{tabular}
\end{table}

\paragraph{R port of the estimators of \citet{yu2024calibration}}
The port uses their MATLAB supplement (doi:10.6084/m9.figshare.24416398). Its backfitting at fixed bandwidths and its exhaustive enumeration reproduce the released MATLAB code to numerical precision (differences below $10^{-10}$ on a test data set; replication repository, Section~\ref{sec:code}). The genetic search itself runs with the R package \texttt{GA} and MATLAB's default population size, so that part of the gap reported in Section~\ref{sec::comparator_mc} may be attributable to the optimizer rather than to the method. One search required $151$ to $545$ generations (median $284$), that is $38$ to $178$ core-hours (median $94$), against about $33$ seconds on one core for \texttt{mgtwr.m}.

\section{Computation times of multiscale spatio-temporal estimators}
\label{app:timing}

This appendix compares the computation time of \texttt{tds\_mgtwr} with that of the other estimators with coefficient-specific spatio-temporal scales considered in this paper, as the sample size $n$ and the number $K$ of varying coefficients increase. Three of them calibrate the $K$ pairs of bandwidths by backfitting, one component at a time---\texttt{tds\_mgtwr}, the Python package \texttt{mgtwr} and the local estimator of \citet{yu2024calibration} (\texttt{mgtwr.m}); the global estimator of \citet{yu2024calibration} (\texttt{mgtwrglobal.m}) searches all pairs jointly with a genetic algorithm, each candidate being a complete backfitting at fixed bandwidths; MGTWR\_2SCALL \citep{wu2026efficient} is a two-step procedure without backfitting, and STVC \citep{murakami2025stvc} relies on an eigenvector basis rather than on local regressions.

\paragraph{Design}
The data follow the balanced panel of Section~\ref{sec::comparator_mc}, the only sampling scheme accepted by all the codes, with linear time: $N$ units on a jittered regular lattice are observed over $T$ periods, with $N\times T=40\times15$, $50\times20$, $100\times25$ and $200\times25$, that is $n=600$, $1{,}000$, $2{,}500$ and $5{,}000$. The number of coefficients, intercept included, is $K=2$, $4$, $8$ or $16$: $\beta_1$ and $\beta_2$ for $K=2$, the four surfaces of the panel design for $K=4$, and, beyond, copies of these four surfaces rotated by $90^\circ$, $180^\circ$ and $270^\circ$ about the center of the study area, time being reversed for the odd rotations, so that the scales to recover remain those of the four original surfaces. Covariates are independent standard normal draws and the noise is scaled as in Section~\ref{sec::comparator_mc}. Each cell is replicated three times (once at $n=5{,}000$) and the median time is reported.

\paragraph{Estimators and settings}
All kernel-based estimators use bi-square kernels with an adaptive spatial bandwidth, so that they share the kernel shape. \texttt{tds\_mgtwr} (\texttt{mgwrsar} version 1.4.2) is run with a spatial bandwidth expressed as a number of neighbors and a fixed temporal bandwidth in periods, with the default settings ($M=20$, OLS initialization) and the convergence tolerance of the Monte Carlo experiments ($10^{-4}$, \ref{B_controls}); the hat matrix required for inference is not computed. The Python package \texttt{mgtwr} (version 2.0.5) is run with an adaptive bi-square kernel, time in periods, a bandwidth searched as a number of observations between $20$ and $n$, the space--time ratio $\tau$ searched in $[0,0.05]$ at six decimals and the default initialization of the package; because of its computing cost, it is not run beyond $K=2$ at $n=5{,}000$, and its longest cells are replicated twice only (marked in Table~\ref{tab:timing_grid}). The local estimator of \citet{yu2024calibration} (\texttt{mgtwr.m}) is run natively in MATLAB with the released files, the exhaustive enumeration of Section~\ref{sec::comparator_mc} replacing the genetic search over the pair of each component. Their global estimator (\texttt{mgtwrglobal.m}) cannot be run to completion on most cells---one search took $38$ to $178$ core-hours on the panel of Section~\ref{sec::comparator_mc}, with $n=2{,}000$ and $K=4$---so its single-core time is extrapolated from the time of one candidate evaluation (note $b$ of Table~\ref{tab:timing_grid}). It is not measured at $n=5{,}000$, nor at $n=2{,}500$ for $K\geq8$, where it exceeds the eight-hour cap by far. MGTWR\_2SCALL is run in MATLAB with the alternating search and the settings of Section~\ref{sec::comparator_mc}. STVC uses up to $500$ spatio-temporal eigenvectors with space--time interactions, the temporal coordinate being the scaled period.

\paragraph{Hardware and measurement}
Every estimation runs on a single core, the BLAS and MATLAB thread counts being set to one, with a cap of eight hours per estimation. The R and Python codes are timed on an Apple M1 Max (R 4.6 linked to the Accelerate BLAS). The MATLAB codes (R2020b) are timed on a Linux computing cluster; their times are converted into M1 equivalents by multiplying them, cell by cell, by the ratio of the times of \texttt{tds\_mgtwr} measured on the two machines with the same package version, which ranges from $0.24$ to $0.43$ (the R port of Section~\ref{sec::comparator_mc} yields similar ratios, from $0.36$ to $0.52$, on the cells with $n\le1{,}000$).

\begin{table}[!htbp]
\caption{Single-core computation times in seconds (median of three replications; at $n=5{,}000$, one replication, two for the Python package) on the balanced panel, by sample size $n$ and number of varying coefficients $K$. Estimation only, without inference, except for the global search, which computes the AICc of the complete model, and thus its projection matrices, as part of calibration. Entries are direct measurements unless the text indicates a hardware conversion or an extrapolation from the generation count. \label{tab:timing_grid}}
\centering
\resizebox{\textwidth}{!}{
\begin{tabular}{r*{24}{r}}
\toprule
 & \multicolumn{16}{c}{Backfitting} & \multicolumn{4}{c}{Two-step} & \multicolumn{4}{c}{Eigenvector-based}\\
\cmidrule(lr){2-17}\cmidrule(lr){18-21}\cmidrule(lr){22-25}
 & \multicolumn{4}{c}{\texttt{tds\_mgtwr}} & \multicolumn{4}{c}{Python \texttt{mgtwr}} & \multicolumn{4}{c}{\shortstack{Yu--Fotheringham\\ local (\texttt{mgtwr.m})$^{a}$}} & \multicolumn{4}{c}{\shortstack{Yu--Fotheringham\\ global (\texttt{mgtwrglobal.m})$^{b}$}} & \multicolumn{4}{c}{MGTWR\_2SCALL$^{a}$} & \multicolumn{4}{c}{STVC}\\
\cmidrule(lr){2-5}\cmidrule(lr){6-9}\cmidrule(lr){10-13}\cmidrule(lr){14-17}\cmidrule(lr){18-21}\cmidrule(lr){22-25}
$n$ & $K{=}2$ & $4$ & $8$ & $16$ & $K{=}2$ & $4$ & $8$ & $16$ & $K{=}2$ & $4$ & $8$ & $16$ & $K{=}2$ & $4$ & $8$ & $16$ & $K{=}2$ & $4$ & $8$ & $16$ & $K{=}2$ & $4$ & $8$ & $16$\\
\midrule
600 & 4.2 & 8.8 & 17 & 22 & 56 & 152 & 523$^{*}$ & 5{,}784 & 19 & 38 & 97 & 360 & 802$^{b}$ & 3{,}759$^{b}$ & 11{,}279$^{b}$ & $>$8\,h$^{\dagger}$ & 1.4 & 2.9 & 6.1 & 17 & 2.2 & 2.5 & 3.3 & 5.6\\
1{,}000 & 8.2 & 15 & 23 & 81 & 117 & 317 & 602$^{*}$ & 2{,}578$^{*}$ & 31 & 132 & 350 & 836 & 2{,}196$^{b}$ & 11{,}611$^{b}$ & $>$8\,h$^{\dagger}$ & $>$8\,h$^{\dagger}$ & 3.7 & 6.0 & 15 & 23 & 3.0 & 3.3 & 4.1 & 7.7\\
2{,}500 & 39 & 76 & 122 & 269 & 358 & 1{,}073$^{*}$ & 2{,}043$^{*}$ & 7{,}018$^{*}$ & 753 & 962 & 3{,}921 & $>$8\,h$^{\dagger}$ & 24{,}143$^{b}$ & $>$8\,h$^{\dagger}$ & $>$8\,h$^{\dagger}$ & $>$8\,h$^{\dagger}$ & 18 & 23 & 59 & 162 & 2.9 & 3.9 & 7.9 & 35\\
5{,}000 & 153 & 257 & 481 & 939 & 972 & -- & -- & -- & 3{,}784 & $>$8\,h$^{\dagger}$ & $>$8\,h$^{\dagger}$ & $>$8\,h$^{\dagger}$ & $>$8\,h$^{\dagger}$ & $>$8\,h$^{\dagger}$ & $>$8\,h$^{\dagger}$ & $>$8\,h$^{\dagger}$ & 67 & 99 & 241 & 548 & 3.4 & 5.4 & 14 & 70\\
\bottomrule
\end{tabular}}
\par\smallskip
\begin{minipage}{\textwidth}\footnotesize
$^{a}$ MATLAB codes timed on a Linux cluster and converted into Apple M1 Max equivalents (see text; at $n=5{,}000$, with the ratio measured at $n=2{,}500$ for the same $K$); local estimator of Yu and Fotheringham with the exhaustive enumeration of Section~\ref{sec::comparator_mc}; MGTWR\_2SCALL with the alternating search. $^{b}$ Global genetic search of Yu and Fotheringham, extrapolated: $P\times$ mean single-core time of one evaluation (a complete backfitting at fixed bandwidths, measured over five generations on the Apple M1 Max with the R port and the \texttt{GA} package) $\times$ median number of generations of the complete searches run on the balanced panel of Section~\ref{sec::comparator_mc} ($284$; $n=2{,}000$, $K=4$), with $P=\min\{\max(20K,40),100\}$; one search, without the restarts suggested by the authors. $^{\dagger}$ Above the eight-hour cap: stopped on the cluster, implied by a run stopped at a smaller $n$ for the same $K$ or, for the global search of Yu and Fotheringham, extrapolated beyond it (not measured at $n=5{,}000$, nor at $n=2{,}500$ for $K\geq8$). --: not run.
$^{*}$ Fewer than three replications.
Converted MATLAB times are approximate hardware-normalized estimates, not direct measurements on the reference machine.
\end{minipage}
\end{table}

\paragraph{Results}
Table~\ref{tab:timing_grid} reports the computation times. \texttt{tds\_mgtwr} takes from $4$ seconds ($n=600$, $K=2$) to about $16$ minutes ($n=5{,}000$, $K=16$); its cost grows roughly linearly with $K$ and less than quadratically with $n$, and it is the fastest backfitting estimator in every cell. The local estimator of \citet{yu2024calibration}, run with an exhaustive enumeration of the integer bandwidth pairs at each update instead of the genetic search of the released code (the enumeration guarantees the optimum of each update, and its time is an upper bound for a search that stops earlier), is $4$ to $32$ times slower and exceeds the eight-hour cap at $n=2{,}500$ with $K=16$; their global search is $190$ to $780$ times slower in the cells where it stays below the cap, and one search would take about $29$ hours at $n=2{,}500$ with $K=4$. The Python package \texttt{mgtwr} is $6$ to $32$ times slower in all cells but one, its cost being driven by the number of sweeps ($52$ to $68$ at $n=600$ with $K=16$, where it takes more than $1.5$ hours against $22$ seconds), which is why it was not run beyond $K=2$ at $n=5{,}000$. MGTWR\_2SCALL, which does not backfit, is $1.3$ to $3.5$ times faster than \texttt{tds\_mgtwr}, and STVC, whose cost depends on the size of its eigenvector basis rather than on $n$, is the fastest overall ($2$ to $70$ seconds). These timings do not compare accuracy, assessed in Sections~\ref{sec::monte_carlo_results} and~\ref{sec::comparator_mc}; with fixed bi-square kernels on both axes, the setting of Section~\ref{sec::comparator_mc}, \texttt{tds\_mgtwr} takes $1.2$ to $2.2$ times longer than with the adaptive spatial bandwidth used here. These times also reflect implementation choices (native routines, QR-based computations, caching, BLAS configuration and exact truncation of compact kernels) and do not isolate the effect of the update order.

\paragraph{Cost of inference}
Table~\ref{tab:computation_times} reports, on the same machine and panel, the single-core times of \texttt{tds\_mgtwr} (\texttt{mgwrsar} 1.4.2) without and with inference, that is, with the projection matrices and the global AICc, for $K=4$ and $K=8$. Inference multiplies the estimation time by $1.4$ to $1.9$, and doubling $K$ by $1.6$ to $2.1$; the largest configuration ($n=5{,}000$, $K=8$, with inference) takes about $15$ minutes. Beyond this size, inference is limited by memory, since it stores an $n\times n$ projection matrix per covariate: at $n=10{,}000$ with $K=4$, the run with inference did not complete on a $64$~GB machine (peak resident memory $38$~GB), whereas estimation alone took $32$ and $59$ minutes for $K=4$ and $K=8$, with about $22$~GB. Estimation and prediction extend further with a neighbor cap: the calibration of the eleven-covariate MGTWR model of Table~\ref{tab:vaucluse} ($40{,}598$ sales, cap of $3{,}000$ neighbors, $M=25$, no projection matrices) took $56$ minutes of wall-clock time with $10$ cores for the candidate evaluations and a peak resident memory of $17$~GB, and predicting the $3{,}638$ withheld sales took $13$ seconds, the covariate screening of Section~\ref{sec:realdata} excluded.

\begin{table}[htbp]
\centering
\caption{Single-core computation times in seconds of \texttt{tds\_mgtwr} (\texttt{mgwrsar} 1.4.2) on an Apple M1 Max, without and with inference (projection matrices and global AICc), by sample size $n$ and number $K$ of varying coefficients: median of three replications (one at $n=5{,}000$) on the balanced panel of this appendix, with the settings of Table~\ref{tab:timing_grid}.}
\label{tab:computation_times}
{
\begin{tabular}{rrrrr}
\toprule
 & \multicolumn{2}{c}{$K=4$} & \multicolumn{2}{c}{$K=8$}\\
\cmidrule(lr){2-3}\cmidrule(lr){4-5}
$n$ & without inference & with inference & without inference & with inference\\
\midrule
600 & 10 & 14 & 18 & 29\\
1{,}000 & 15 & 23 & 26 & 39\\
2{,}500 & 76 & 108 & 125 & 184\\
5{,}000 & 255 & 451 & 483 & 907\\
\bottomrule
\end{tabular}}
\end{table}

\section{Estimation and specification settings of the applications}
\label{B_apps}

\paragraph{Temporal kernels}
Symmetric rather than past-only temporal kernels \citep{zhang2021multiscale} are retained: observation dates rarely coincide with infection onset in disease monitoring, and variable delays separate decision, transaction and recorded sale in housing markets, so that limited borrowing from near-future observations buffers this timing misalignment.

\paragraph{Sharpening exponent}
The exponent $\gamma$ of the prediction kernels (Section~\ref{subsec:prediction}) is chosen by an internal validation that reproduces the forecasting design within the training sample, the test observations playing no part. For beet yellows severity, the last ten days of the in-sample window of every training season are withheld and predicted from the earlier observations. For house prices, the sales of 2021 are predicted from those of 2007--2020. The exponent minimizing the validation RMSE on the grid $\{1, 2, 4, 8, 12, 16, 24, 32, 48, 64\}$ is retained, the grid stopping at $32$ for house prices (Table~\ref{tab:gamma}). This choice matters little: the validation curves are flat beyond $\gamma=8$, the test error moves by less than $0.001$ between a large retained value and $\gamma=24$ (between $\gamma=20$ and $\gamma=32$ for house prices), and by up to $0.01$ in the four cells where $\gamma=4$ is retained.

\begin{table}[htbp]
\centering
\caption{Sharpening exponent $\gamma$ retained by the internal validation.}
\label{tab:gamma}
\footnotesize
\begin{tabular}{lccccc}
\toprule
 & \multicolumn{4}{c}{Beet yellows severity} & House prices \\
\cmidrule(lr){2-5}
Model & 2020 & 2021 & 2022 & 2023 & 2022 \\
\midrule
GTWR & $32$ & $4$ & $4$ & $4$ & $32$ \\
\texttt{tds\_mgtwr} & $64$ & $64$ & $64$ & $4$ & $24$ \\
\bottomrule
\end{tabular}
\end{table}

\paragraph{Global, machine-learning and eigenvector benchmarks}
Table~\ref{tab:bench_settings} gives the settings. MARS uses an additive basis because a degree-two basis fitted on beet yellows severity extrapolates its season by past-severity interaction linearly beyond the training range of the 2020 outbreak and doubles the 2020 RMSE. The tree depth of XGBoost is chosen by a validation that reproduces the forecasting design. For beet yellows severity, the late-season observations (days $201$--$230$) of each other season are predicted in turn, and for house prices, the 2021 sales are predicted from the 2007--2020 sales. Folds and subsampling being random, its scores are means over $20$ seeds, with a standard deviation of the RMSE between $0.002$ (2022) and $0.028$ (2020) for beet yellows severity and of $0.0008$ for house prices. STVC imposes no cyclic structure, in line with the non-cyclic kernels of GTWR and MGTWR, and each covariate has its own pattern of spatio-temporal variation through the interactions between eigenvectors and covariates. For house prices, its stepwise search adds the distances to the post office, to the maternity unit and to public transport to the set of the kernel-based models and drops the distance to the hospital.

\begin{table}[htbp]
\centering
\caption{Settings of the benchmarks of the applications.}
\label{tab:bench_settings}
\footnotesize
\begin{tabular}{@{}>{\raggedright\arraybackslash}p{1.6cm} >{\raggedright\arraybackslash}p{6.0cm} >{\raggedright\arraybackslash}p{9.2cm}@{}}
\toprule
Model & Covariates & Settings \\
\midrule
OLS & forward--backward stepwise on the AIC (\texttt{stepAIC}, R package \texttt{MASS}), from all candidates, except the time index and the month for house prices & -- \\
MARS & all candidates and the coordinates & \texttt{earth}: at most $200$ candidate terms, additive basis, default generalized cross-validation pruning \\
XGBoost & all candidates and the coordinates & learning rate $0.1$, \texttt{min\_child\_weight}$=1$; subsampling of rows and columns $0.85$ (beet yellows severity), none (house prices); rounds by five-fold cross-validation with early stopping, then refit on the whole training sample; depth among $2$ to $9$: $9$ for 2020, $4$ for 2021--2023, $6$ for house prices \\
STVC & forward--backward stepwise on the AIC of \texttt{spmoran}, from all candidates except the time index and the day of year (beet yellows severity) or from the set of the kernel-based models (house prices) & up to $500$ spatio-temporal eigenvectors, with the reluctant-interaction mechanism; temporal basis: day of year and linear time index (beet yellows severity), month index and year of sale (house prices) \\
\bottomrule
\end{tabular}
\end{table}

\paragraph{Covariate selection and local multicollinearity}
Table~\ref{tab:selection} gives the selection of the kernel-based models, which uses the training sample only. Its two-stage design, a collinearity screening followed by an information-criterion selection, follows from the controlled simulations of \citet{kang2025scale}: coefficient-specific bandwidths make MGWR more robust than GWR to local multicollinearity, but very strong collinearity, especially combined with spatially autocorrelated covariates, can still distort bandwidth selection, create false local variation or oversmooth genuine local effects. The screening proved necessary on beet yellows severity, where meteorological, landscape and vector-related covariates are strongly correlated within restricted neighborhoods, and much less so on house prices, where covariates are more weakly correlated and local samples larger. Concurvity applies only to \texttt{tds\_mgtwr}, whose components can be explained by one another, and complements the candidate pre-selection of Section~\ref{sec:algo}, which acts on the bandwidths rather than on the covariates. A stepwise search depends on where it starts, so that running the beet yellows selection twice, with the two temporal trends of Table~\ref{tab:selection}, reduces the risk of ending in a poor solution: in several cells, the run with the imposed trend reached the lower AICc. Since each estimator is selected by its own AICc on beet yellows severity, the comparisons of Table~\ref{tab:beet_all} bear on complete pipelines rather than on a common specification. All sets were fixed before the out-of-sample evaluation, and the resulting specifications are those of Tables~\ref{tab:covars_2020} and~\ref{tab:vaucluse_bw}.

\begin{table}[htbp]
\centering
\caption{Covariate selection of the kernel-based models.}
\label{tab:selection}
\footnotesize
\begin{tabular}{@{}>{\raggedright\arraybackslash}p{2.3cm} >{\raggedright\arraybackslash}p{8.2cm} >{\raggedright\arraybackslash}p{6.3cm}@{}}
\toprule
Step & Beet yellows severity & House prices \\
\midrule
Candidates & $15$ (Table~S1), including the time index, the day of year and the state variables \texttt{S\_log} and \texttt{S\_box} & $18$ (Section~\ref{subsec:datasets}) \\
Screening & concurvity, \texttt{tds\_mgtwr} only: a covariate is removed when its component is explained at more than $80\%$ ($R^2>0.8$) by the other components & local collinearity screening: month, quarter, time index and distances to the post office and to the maternity unit removed \\
Starting set & core of $9$: the temporal trend, \texttt{DD\_Qi}, \texttt{t\_q.M4}, \texttt{etp\_q.M3}, \texttt{meadows\_5km}, \texttt{dist\_PG}, \texttt{seed\_treatments}, \texttt{aphids\_pct} and \texttt{S\_log} & the $13$ remaining covariates \\
Search & backward: at each round, the removal that lowers the AICc most, as long as it does not raise the AICc by $10$ points or more; forward: candidates by decreasing absolute correlation with the residuals, the first that lowers the AICc by at least $10$ points being added, with a concurvity below $0.8$ for \texttt{tds\_mgtwr}, until none qualifies & forward--backward stepwise, the bandwidth being re-optimized for each candidate set \\
Criterion & AICc of the estimator: that of the bandwidth pair for GTWR, the global AICc of \citet{yu2020inference} for \texttt{tds\_mgtwr} & AICc of the single-bandwidth GWR for all kernel-based models, the global AICc not being computed at this sample size (\ref{app:timing}) \\
Identifiability floors & \texttt{tds\_mgtwr}: candidate pre-selection (Section~\ref{sec:algo}); GTWR: the same local-variance rule on the common spatial bandwidth, at least $20$ neighbors; a $14$-day temporal floor for both & \texttt{tds\_mgtwr}: candidate pre-selection (Section~\ref{sec:algo}); GTWR: search from $20$ neighbors and $3$ months \\
Temporal trend & selection run twice, with the day of year in the core and the time index as a candidate, then with the time index imposed and the day of year excluded; the lower AICc is retained, the imposed trend being kept when the gap is below $10$ points & year of sale among the candidates, and retained \\
Result & season-specific sets (Table~\ref{tab:covars_2020}) & $11$ covariates: share of pre-1945 dwellings and distance to public transport removed \\
\bottomrule
\end{tabular}
\end{table}

\begin{table}[htbp]
\centering
\caption{Covariates included and selected bandwidths by model and year for GTWR and \texttt{tds\_mgtwr} with the Beet Yellows Severity dataset, for the models of Table~\ref{tab:beet_all}. Spatial bandwidths $h_s$ are numbers of nearest neighbors (adaptive kernels) and temporal bandwidths $h_t$ are in days. A spatial bandwidth equal to the training sample size ($2{,}667$, $3{,}030$, $2{,}914$ and $3{,}001$ from 2020 to 2023) or a temporal bandwidth equal to the largest temporal gap ($1{,}598$ days, $1{,}556$ in 2023) is a boundary choice: no weighting along that axis. Empty cell: covariate not in the model.}
\label{tab:covars_2020}

\resizebox{1\linewidth}{!}{
\begin{tabular}{lcccc lcccc lcccc lcccc}
\toprule
 & \multicolumn{4}{c}{\textbf{2020}} 
 & \multicolumn{4}{c}{\textbf{2021}} 
 & \multicolumn{4}{c}{\textbf{2022}} 
 & \multicolumn{4}{c}{\textbf{2023}} \\
\cmidrule(lr){2-5}\cmidrule(lr){6-9}\cmidrule(lr){10-13}\cmidrule(lr){14-17}
\textbf{Variable}
& \multicolumn{2}{c}{\textbf{GTWR}}
& \multicolumn{2}{c}{\textbf{tds\_mgtwr}}
& \multicolumn{2}{c}{\textbf{GTWR}}
& \multicolumn{2}{c}{\textbf{tds\_mgtwr}}
& \multicolumn{2}{c}{\textbf{GTWR}}
& \multicolumn{2}{c}{\textbf{tds\_mgtwr}}
& \multicolumn{2}{c}{\textbf{GTWR}}
& \multicolumn{2}{c}{\textbf{tds\_mgtwr}} \\
& $h_s$ & $h_t$ & $h_s$ & $h_t$
& $h_s$ & $h_t$ & $h_s$ & $h_t$
& $h_s$ & $h_t$ & $h_s$ & $h_t$
& $h_s$ & $h_t$ & $h_s$ & $h_t$ \\
\midrule

Intercept
& 35 & 83 & 14 & 1598
& 39 & 178 & 6 & 187
& 35 & 196 & 6 & 839
& 39 & 193 & 11 & 1556 \\

time
&  &  &  & 
& 39 & 178 & 78 & 1598
& 35 & 196 & 44 & 1598
& 39 & 193 & 77 & 1556 \\

day
& 35 & 83 & 8 & 150
&  &  &  & 
&  &  &  & 
&  &  &  &  \\

DD\_Qi
& 35 & 83 & 73 & 1598
& 39 & 178 &  & 
& 35 & 196 & 25 & 1598
& 39 & 193 & 14 & 1556 \\

meadows\_5km
& 35 & 83 & 291 & 355
& 39 & 178 &  & 
& 35 & 196 & 310 & 677
& 39 & 193 &  &  \\

t\_q.M4
&  &  &  & 
&  &  &  & 
&  &  & 718 & 22
&  &  & 3001 & 183 \\

Tm\_spring
&  &  &  & 
& 39 & 178 & 78 & 839
& 35 & 196 &  & 
& 39 & 193 & 33 & 183 \\

tsup\_h\_q.M4
&  &  &  & 
&  &  & 3030 & 231
&  &  &  & 
&  &  &  &  \\

etp\_q.M3
&  &  & 221 & 79
&  &  & 982 & 18
&  &  & 310 & 121
&  &  & 239 & 97 \\

preliq\_q.M1
&  &  & 168 & 150
&  &  &  & 
&  &  & 310 & 1598
&  &  &  &  \\

seed\_treatments
& 35 & 83 & 73 & 1598
& 39 & 178 & 78 & 1598
&  &  & 234 & 1598
& 39 & 193 & 77 & 1556 \\

dist\_PG
& 35 & 83 & 384 & 18
& 39 & 178 & 422 & 18
& 35 & 196 & 410 & 98
& 39 & 193 & 419 & 17 \\

aphids\_pct
& 35 & 83 & 73 & 440
& 39 & 178 & 44 & 440
& 35 & 196 & 25 & 1598
& 39 & 193 & 25 & 348 \\

S\_log
&  &  & 18 & 287
& 39 & 178 & 14 & 1598
& 35 & 196 & 11 & 1598
& 39 & 193 & 11 & 1556 \\

S\_box
&  &  &  & 
&  &  & 103 & 287
&  &  & 76 & 546
&  &  &  &  \\
\bottomrule
\end{tabular}}
\end{table}

\section{Standard errors in multiscale backfitting: local approximation versus exact method}
\label{B_inf}

\paragraph{Comparing local and backfitting-based standard errors in multiscale frameworks}

To evaluate the inferential risks associated with standard univariate approximations in multiscale GWR, Table~\ref{tab:inference_comparison_final} presents a comparative analysis of mean standard errors (SE) and the spatial distribution of $t$-statistics, using the 2020 training sample of the Beet Yellows Severity dataset as an illustrative example. We contrast the exact method, derived from the covariate-specific projection matrices of the backfitting \citep{yu2020inference}, whose sum gives the effective number of parameters and thus accounts for the loss of degrees of freedom induced by the iterative fitting, with a local approximation that takes, for each component, the variance of its final univariate GWR fit, as if this fit were an ordinary GWR of the partial residual.

For clarity, the term “exact” is used here in a restricted sense, referring to inference computed within the linearized backfitting approximation underlying the multiscale estimator. The reference inference procedure relies on a global residual variance estimator,
\[
\hat{\sigma}^2 \;=\; \frac{\sum_{i=1}^{n} e_i^2}{\,n - \mathrm{tr}(\mathbf{S}) - p_f\,},
\]
where $\mathrm{tr}(\mathbf{S})$ denotes the effective degrees of freedom induced by the fitted multi-scale smoother and $p_f$ is the number of fixed covariates. Standard errors for the spatially and temporally varying coefficients are obtained as:
\[
\widehat{SE}\{\hat{\beta}_k(\cdot)\}_i \;=\; \sqrt{\hat{\sigma}^2 \, v_{ik}},
\]
where $v_{ik} = \sum_{j} R_{k}(i,j)^2 / x_{ik}^2$, $\mathbf{R}_k$ being the $n \times n$ matrix that maps the response to the fitted contribution of covariate $k$ at the returned state, $x_{ik}\hat{\beta}_k(i) = (\mathbf{R}_k \mathbf{y})_i$, so that $\mathbf{S} = \sum_k \mathbf{R}_k$. The local approximation replaces $\mathbf{R}_k$ by the smoother $\mathbf{S}_k$ of the final univariate fit of component $k$, that is $v_{ik}^{\mathrm{loc}} = \sum_{j} S_{k}(i,j)^2 / x_{ik}^2$, and $\hat{\sigma}^2$ by the local residual variance of this fit. It thus treats the other components as known: it ignores the dependence that the backfitting creates between components, and the collinearity between their fitted contributions, which $\mathbf{R}_k$ carries, and its local residual variance is on average smaller than $\hat{\sigma}^2$.

On this example, the local approximation yields smaller standard errors than the exact method: the mean standard error is $1.4$ to $4.8$ times smaller depending on the coefficient ($1.4$ for \texttt{preliq\_q.M1}, $4.8$ for the intercept). The share of locations flagged as significant changes accordingly. For \texttt{meadows\_5km}, it falls at the $1\%$ level from $65.02\%$ under the local approximation to $1.35\%$ with the exact standard errors, and to $0\%$ after the FDR adjustment of \citet{Yekutieli2001}, which controls for multiple testing under arbitrary dependence and is applied separately to each covariate, over its $n=2{,}667$ local tests (one family per mapped coefficient, not one across all coefficients); the corresponding shares are $92.99$, $32.77$ and $19.54\%$ for the intercept, $42.37$, $26.02$ and $15.52\%$ for \texttt{preliq\_q.M1}, and $94.15$, $35.62$ and $31.80\%$ for \texttt{seed\_treatments}. Other covariates keep a substantial localized signal under the exact method and the FDR adjustment, such as \texttt{dist\_PG} ($92.28\%$), \texttt{day} ($93.33\%$), \texttt{DD\_Qi} ($42.89\%$) and \texttt{etp\_q.M3} ($42.11\%$). Each column reports its own variance calculation and multiple-testing treatment: the contrast between the local and the FDR-adjusted exact columns combines both. These differences motivate caution in significance mapping with local approximations; on these empirical data alone, however, they do not measure type-I error rates or coverage, which would require a simulation with known coefficients. Both variance calculations are conditional on the selected covariates, bandwidths, update path and retained sweep, and assume independent homoskedastic disturbances; neither accounts for the adaptive selection that precedes them.

\begin{table}[htbp]
\centering
\caption{Comparison of inferential results: Local approximation vs. Exact method (Raw and FDR corrected); MGTWR model of the 2020 season retained in Table~\ref{tab:beet_all}}
\label{tab:inference_comparison_final}
\resizebox{\linewidth}{!}{
\begin{tabular}{lccccccccccc}
\toprule
 & \multicolumn{4}{c}{Local Approximation ($\mathbf{S}_k$)} & \multicolumn{7}{c}{Exact Method ($\mathbf{R}_k$)} \\
\cmidrule(lr){2-5}\cmidrule(lr){6-12}
 & Mean & \multicolumn{3}{c}{\% Significant} & Mean & \multicolumn{3}{c}{\% Significant} & \multicolumn{3}{c}{\% Sign. FDR adjusted} \\
\cmidrule(lr){3-5}\cmidrule(lr){7-9}\cmidrule(lr){10-12}
Variables & S.E. & 10\% & 5\% & 1\% & S.E. & 10\% & 5\% & 1\% & 10\% & 5\% & 1\% \\
\midrule
Intercept       & 0.0586 & 94.98 & 94.56 & 92.99 & 0.2800 & 70.60 & 54.22 & 32.77 & 25.65 & 23.73 & 19.54 \\
day             & $7.6{\times}10^{-4}$ & 97.11 & 96.33 & 95.24 & 0.0018 & 97.68 & 96.93 & 95.13 & 95.24 & 95.05 & 93.33 \\
DD\_Qi          & 0.0269 & 93.51 & 88.08 & 85.49 & 0.0687 & 66.03 & 64.30 & 53.58 & 49.49 & 46.72 & 42.89 \\
etp\_q.M3       & 0.0354 & 89.13 & 87.89 & 86.05 & 0.0952 & 65.02 & 59.96 & 48.74 & 44.77 & 43.27 & 42.11 \\
meadows\_5km   & 0.0193 & 75.59 & 73.30 & 65.02 & 0.0411 & 35.81 & 19.87 & 1.35 & 0.00 & 0.00 & 0.00 \\
dist\_PG        & 0.0372 & 97.64 & 97.41 & 96.93 & 0.1238 & 93.81 & 92.61 & 92.46 & 92.46 & 92.43 & 92.28 \\
seed\_treatments              & 0.0264 & 95.05 & 94.79 & 94.15 & 0.0627 & 48.29 & 39.15 & 35.62 & 34.95 & 33.52 & 31.80 \\
aphids\_pct     & 0.0365 & 75.59 & 70.83 & 64.00 & 0.0614 & 47.81 & 42.52 & 38.36 & 37.98 & 37.46 & 36.71 \\
S\_log          & 0.0736 & 95.69 & 94.86 & 90.66 & 0.1340 & 61.49 & 55.91 & 44.84 & 40.87 & 38.02 & 33.56 \\
preliq\_q.M1    & 0.0426 & 61.90 & 51.74 & 42.37 & 0.0610 & 38.47 & 32.40 & 26.02 & 20.70 & 19.20 & 15.52 \\
\bottomrule
\end{tabular}}
\vspace{0.6em}

\footnotesize\emph{Note}: $n=2667$. \% Significant report the proportion of locations where $H_0:\beta_k(\cdot)=0$ is rejected. FDR: False Discovery Rate correction \citep{Yekutieli2001} applied to the exact method's p-values, separately for each covariate over its $n$ local tests.
\end{table}

\clearpage
\section*{Supplementary Material}
\addcontentsline{toc}{section}{Supplementary Material}
\setcounter{table}{0}
\renewcommand{\thetable}{S\arabic{table}}
\renewcommand{\theHtable}{S\arabic{table}}

\subsection*{S1. Beet yellows severity: candidate covariates}\label{S_beet_covars}

Table~\ref{tab:S_beet_dict} lists the $15$ candidate covariates of the beet yellows severity application (Section~\ref{subsec:datasets}), with the names used in the result tables. Lagged meteorological summaries are computed over calendar months preceding the observation date (``month $m$ preceding'' denotes the $m$-th 30-day period before the visit); the winter and spring summaries are fixed within a season. The two regional indicators of past severity are common to all fields at a given date and use only severities observed at least $15$ days earlier. In the retrospective design of the article, every covariate of a withheld observation is therefore computed from information dated before that observation (weather lags, winter and spring summaries, past severity) or recorded at its visit (aphid counts, seed treatment); the aphid count is a contemporaneous covariate, not a forecast input.

\begin{table}[htbp]
\centering
\caption{Candidate covariates of the beet yellows severity application.}
\label{tab:S_beet_dict}
\small
\begin{tabular}{>{\ttfamily}l l p{9.2cm}}
\toprule
\textrm{Name} & Family & Definition \\
\midrule
DD\_Qi & meteorology (winter) & degree days between January 1 and February 14 \\
Tm\_spring & meteorology (spring) & mean temperature between February 1 and July 1 ($^\circ$C) \\
t\_q.M4 & meteorology (lagged) & mean temperature during the fourth month preceding the observation ($^\circ$C) \\
tsup\_h\_q.M4 & meteorology (lagged) & mean maximum temperature during the fourth month preceding the observation ($^\circ$C) \\
etp\_q.M3 & meteorology (lagged) & potential evapotranspiration during the third month preceding the observation (mm) \\
preliq\_q.M1 & meteorology (lagged) & liquid precipitation during the first month preceding the observation (mm) \\
q\_q.M4 & meteorology (lagged) & specific humidity during the fourth month preceding the observation (g/kg) \\
meadows\_5km & land cover & share of grassland within 5~km of the field \\
dist\_PG & landscape context & distance to the nearest sugar-beet seed-production field \\
seed\_treatments & management & seed treatment (binary) \\
aphids\_pct & biotic pressure & percentage of plants with aphids among $20$ sampled plants at the observation date, smoothed over $12$~km and the previous $12$ days \\
S\_log & past severity (regional) & logarithm of one plus the sum of the severities observed in the same season at least $15$ days before the observation date \\
S\_box & past severity (regional) & same sum restricted to severities observed between $15$ and $30$ days before the observation date \\
day & temporal index & day of year \\
time & temporal index & day number over the whole series (2019--2023) \\
\bottomrule
\end{tabular}
\end{table}

\end{document}